\documentclass[pdflatex,sn-mathphys]{sn-jnl}

\jyear{2022}%

\normalbaroutside 
\usepackage[acronym]{glossaries} 
\usepackage[caption=false]{subfig} 
\usepackage{adjustbox} 
\usepackage[multiple]{footmisc}
\usepackage{refcount}
\usepackage{amssymb}
\usepackage{pifont}
\newcommand{\cmark}{\ding{51}}%
\newcommand{\xmark}{\ding{55}}%
\usepackage{makecell}

\begin{document}



%

\title[ControlPULP: A RISC-V On-Chip Parallel Power Controller for HPC]{ControlPULP: A RISC-V On-Chip Parallel Power Controller for Many-Core HPC Processors with FPGA-Based Hardware-In-The-Loop Power and Thermal Emulation}



\author*[1]{\fnm{Alessandro} \sur{Ottaviano}}\email{aottaviano@iis.ee.ethz.ch}

\author[1]{\fnm{Robert} \sur{Balas}}\email{balasr@iis.ee.ethz.ch}

\author[2]{\fnm{Giovanni} \sur{Bambini}}\email{giovanni.bambini2@unibo.it}

\author[2]{\fnm{Antonio} \sur{del Vecchio}}\email{antonio.delvecchio5@unibo.it}

\author[2]{\fnm{Maicol} \sur{Ciani}}\email{maicol.ciani@unibo.it}

\author[2]{\fnm{Davide} \sur{Rossi}}\email{davide.rossi@unibo.it}

\author[1,2]{\fnm{Luca} \sur{Benini}}\email{lbenini@iis.ee.ethz.ch}

\author[2]{\fnm{Andrea} \sur{Bartolini}}\email{a.bartolini@unibo.it}

\affil*[1]{\orgdiv{Integrated Systems Laboratory}, \orgname{ETH Zürich}, \orgaddress{\street{Gloriastrasse 35}, \city{Zürich}, \postcode{8092}, \country{Switzerland}}}

\affil[2]{\orgdiv{DEI}, \orgname{University of Bologna}, \orgaddress{\street{Viale Del Risorgimento 2}, \city{Bologna}, \postcode{40136}, \country{Italy}}}

\ifx\showrebuttal\undefined
    \newcommand{\revision}[1]{{#1}}
\else
    \newcommand{\revision}[1]{{\textcolor{red}{#1}}}
\fi


\abstract{
High-Performance Computing (HPC) processors are nowadays integrated Cyber-Physical Systems demanding complex and high-bandwidth closed-loop power and thermal control strategies. To efficiently satisfy real-time multi-input multi-output (MIMO) optimal power requirements, high-end processors integrate an on-die power controller system (PCS).

While traditional PCSs are based on a simple microcontroller (MCU)-class core, more scalable and flexible PCS architectures are required to support advanced MIMO control algorithms for managing the ever-increasing number of cores, power states, and process, voltage, and temperature variability.

This paper presents ControlPULP, an open-source, HW/SW RISC-V parallel PCS platform consisting of a single-core MCU with fast interrupt handling coupled with a scalable multi-core programmable cluster accelerator and a specialized DMA engine for the parallel acceleration of real-time power management policies. ControlPULP relies on FreeRTOS to schedule a reactive power control firmware (PCF) application layer.

We demonstrate ControlPULP in a power management use-case targeting a next-generation 72-core HPC processor.
We first show that the multi-core cluster accelerates the PCF, achieving 4.9x speedup compared to single-core execution, enabling more advanced power management algorithms within the control hyper-period at a shallow area overhead, about 0.1\% the area of a modern HPC CPU die.
We then assess the PCS and PCF by designing an FPGA-based, closed-loop emulation framework that leverages the heterogeneous SoCs paradigm, achieving DVFS tracking with a mean deviation within 3\% the plant's thermal design power (TDP) against a software-equivalent model-in-the-loop approach.
Finally, we show that the proposed PCF compares favorably with an industry-grade control algorithm under computational-intensive workloads.
}



\newacronym{dtm}{DTM}{dynamic thermal management}
\newacronym{stm}{STM}{static thermal management}
\newacronym{hw}{HW}{hardware}
\newacronym{sw}{SW}{software}
\newacronym{ca}{CA}{command/address}
\newacronym{ip}{IP}{intellectual property}
\newacronym{ddr}{DDR}{double data rate}
\newacronym{lpddr}{LPDDR}{low-power double data rate}
\newacronym{rpc}{RPC}{reduced pin count}
\newacronym{dma}{DMA}{direct memory access}
\newacronym{axi}{AXI}{Advanced eXtensible Interface}
\newacronym{dram}{DRAM}{dynamic random access memory}
\newacronym[firstplural=static random access memories (SRAMs)]{sram}{SRAM}{static random access memory}
\newacronym{edram}{eDRAM}{embedded DRAM}
\newacronym[firstplural=systems on chip (SoCs)]{soc}{SoC}{system on chip}
\newacronym{mpsoc}{MPSoC}{multi-processor system on chip}
\newacronym{hesoc}{HeSoC}{heterogeneous system on chip}
\newacronym{sip}{SiP}{system in package}
\newacronym{fpga}{FPGA}{field-programmable gate array}
\newacronym{asic}{ASIC}{application-specific integrated circuit}
\newacronym{phy}{PHY}{physical layer}
\newacronym{ml}{ML}{machine learning}
\newacronym{iot}{IoT}{internet of things}
\newacronym{foss}{FOSS}{free and open source}
\newacronym{cmos}{CMOS}{complementary metal-oxide-semiconductor}
\newacronym{sut}{SUT}{system under test}
\newacronym{isut}{ISUT}{integrated system under test}
\newacronym{rtl}{RTL}{register transfer level}
\newacronym{hil}{HIL}{hardware in the loop}
\newacronym{pil}{PIL}{processor in the loop}
\newacronym{fil}{FIL}{FPGA in the loop}
\newacronym{mil}{MIL}{model in the loop}
\newacronym{sil}{SIL}{software in the loop}
\newacronym{hpc}{HPC}{high performance computing}
\newacronym{mcu}{MCU}{microcontroller unit}
\newacronym{fub}{FUB}{functional unit block}
\newacronym{ecu}{ECU}{electronic control unit}
\newacronym{dcu}{DCU}{domain control unit}
\newacronym{adas}{ADAS}{advanced driver-assistance system}
\newacronym{fame}{FAME}{FPGA Architecture Model Execution}
\newacronym{pl}{PL}{Programmable Logic}
\newacronym{ps}{PS}{Processing System}
\newacronym{apu}{APU}{Application Processing Unit}
\newacronym{ocm}{OCM}{on-chip memory}
\newacronym{pcs}{PCS}{power controller system}
\newacronym{pcf}{PCF}{power control firmware}
\newacronym{plic}{PLIC}{platform-level interrupt controller}
\newacronym{pmca}{PMCA}{programmable many-core accelerator}
\newacronym{bram}{BRAM}{block RAM}
\newacronym{lut}{LUT}{look-up table}
\newacronym{ff}{FF}{flip-flop}
\newacronym{fsbl}{FSBL}{First Stage BootLoader}
\newacronym{pvt}{PVT}{Process, Voltage, Temperature}
\newacronym{hls}{HLS}{high-level synthesis}
\newacronym{mqtt}{MQTT}{Message Queuing Telemetry Transport}
\newacronym{cots}{COTS}{commercial off-the-shelf}
\newacronym{cpu}{CPU}{central processing unit}
\newacronym{gpu}{GPU}{graphic processing unit}
\newacronym{ibmocc}{IBM OCC}{IBM on-chip controller}
\newacronym{clic}{CLIC}{core-local interrupt controller}
\newacronym{clint}{CLINT}{core-local interruptor}
\newacronym{scmi}{SCMI}{System Control and Management Interface}
\newacronym{os}{OS}{Operating System}
\newacronym{mimo}{MIMO}{multiple-input multiple-output}
\newacronym{bmc}{BMC}{Baseboard Management Controller}
\newacronym{qos}{QoS}{quality of service}
\newacronym{tdp}{TPD}{thermal design power}
\newacronym{dvfs}{DVFS}{dynamic voltage and frequency scaling}
\newacronym{dfs}{DFS}{dynamic frequency scaling}
\newacronym{dvs}{DVS}{dynamic voltage scaling}
\newacronym{rtu}{RTU}{Real Time Unit}
\newacronym{pe}{PE}{processing element}
\newacronym{noc}{NoC}{network on chips}
\newacronym{pid}{PID}{proportional integral derivative}
\newacronym{sota}{SOTA}{state-of-the-art}
\newacronym{fpu}{FPU}{floating point unit}
\newacronym{pcu}{PCU}{Power Control Unit}
\newacronym{scp}{SCP}{System Control Processor}
\newacronym{mcp}{MCP}{Manageability Control Processor}
\newacronym{occ}{OCC}{On-Chip Controller}
\newacronym{smu}{SMU}{System Management Unit}
\newacronym{ap}{AP}{application-class processor}
\newacronym{vrm}{VRM}{voltage regulator module}
\newacronym{pfct}{PFCT}{periodic frequency control task}
\newacronym{pvct}{PVCT}{periodic voltage control task}
\newacronym{ipc}{IPC}{instructions per cycle}
\newacronym{simd}{SIMD}{single instruction, multiple data}
\newacronym{mctp}{MCTP}{Management Component Transport Protocol}
\newacronym{pmbus}{PMBUS}{Power Management Bus}
\newacronym{avsbus}{AVSBUS}{Adaptive Voltage Scaling}
\newacronym{pldm}{PLDM}{Platform Level Data Model}
\newacronym{rtos}{RTOS}{real-time OS}
\newacronym{hlc}{HLC}{high-level controller}
\newacronym{llc}{LLC}{low-level controller}
\newacronym{shv}{SHV}{selective hardware vectoring}
\newacronym{api}{API}{application programming interface}
\newacronym{hal}{HAL}{hardware abstraction layer}
\newacronym{smp}{SMP}{symmetric multiprocessing}
\newacronym{pll}{PLL}{phase locked loop}




\maketitle

\section{Introduction}

After the end of Dennard's scaling, the increase in power density has become an undesired but unavoidable collateral effect of the performance gain obtained with integrated systems' technological scaling. 
An increase in power density has multiple adverse effects, which are collectively referred to as the \textit{power wall}: component's lifetime shortening, electromigration, dielectric breakdown due to thermal hot spots and sharp thermal gradients, and degraded operating speed due to leakage current exponentially increasing with temperature. 
This trend has made the \glspl{pe} at the heart of computing nodes energy, power, and thermally constrained~\cite{Leisersoneaam9744}.

Two approaches have been adopted to mitigate the \textit{power wall} at the system level~\cite{Tilli2022}: \gls{stm} and \gls{dtm} techniques. The former allows increasing the \gls{tdp} sustained by the chip with a tailored design of heat sinks, fans, and liquid cooling. However, \gls{stm} strategies incur increasingly unsustainable costs when over-designed to remove heat in the worst-case conditions for today’s \gls{hpc} processors. Consequently, \gls{dtm} techniques have become more and more crucial to bound the operating temperature with \textit{run-time active control}, for example, by exploiting \gls{pvt} sensors along with \gls{dvfs}, thread migration/scheduling, throttling, and clock gating. Hence, standard cooling systems can be designed to handle the average case, leaving the management of peaks to active control.

Modern high-performance processors feature many cores integrated into a single silicon die. Recent notable examples are AWS Graviton 3 (64 Arm Neoverse V1 cores)~\cite{GRAVITON}, Intel Raptor Lake (24 cores, 32 threads)~\cite{RAPTORLAKE}, AMD Epyc 7004 Genoa (up to 96 Zen 4 cores)~\cite{AMDGENOA}, SiPearl Rhea Processor (72 Arm Neoverse V1 Zeus cores)~\cite{SIPEARL}, Ampere Altra Max (128 Arm Neoverse N1), and the NVIDIA Grace \gls{cpu} (144 Arm Neoverse V2 cores).
Their application workload requires runtime dynamic trade-off between maximum performance (fastest operating point~\cite{Cesarini}) in \gls{cpu}-bound execution phases and energy efficiency in memory-bound execution phases (energy-aware \gls{cpu}~\cite{SANDYBRIDGE}).

While software-centric advanced \gls{dtm} policies have been proposed~\cite{BARTOLINI_MPC, BENEVENTI_ONLINE, Tilli2022}, they mainly execute on the \gls{cpu}'s \glspl{ap}, playing the role of \glspl{hlc} governors. 
Nevertheless, in recent years it has become clear the trend of abstracting power and system management tasks away from the \glspl{ap}~
\footnote{\label{scp_open}\url{https://github.com/Arm-software/SCP-firmware}}~\cite{SCMIWHITE} towards control systems that are closer to the controlled hardware components and can guarantee faster, and more precise control actions, namely \glspl{llc}.

Modern processors integrate on-die \glspl{llc}~\cite{PERCOREPSTATE} in the \textit{uncore} domain, referred to as \glspl{pcs}, as dedicated embedded hardware resources, co-designed with a \gls{pcf} implementing complex \gls{mimo} power management policies.
Advanced \gls{dtm} involves embedding and interleaving a plurality of activities in the \gls{pcs}, namely (i) dynamic control of the \gls{cpu} power consumption with short time constants~\cite{HYPERPERIOD}, required to prevent thermal hazards and to meet the \gls{tdp} limit (power capping~\cite{PWRMM_TECHNIQUES}), (ii) real-time interaction with commands provided by on-die (\gls{os} - power management interfaces and on-chip sensors) and off-die (\gls{bmc}, \glspl{vrm}) units and (iii) dynamic power budget allocation between general-purpose (\glspl{cpu}) and other integrated subsystems, such as \glspl{gpu}~\cite{SANDYBRIDGE}.

Existing on-die \glspl{pcs} share a similar design structure. They feature an embedded single-core \gls{mcu}~\footnotemark[\getrefnumber{scp_open}] supported by dedicated hardware state machines~\cite{SANDYBRIDGE} or more generic accelerators~\cite{ibm_occ}.
The hardware typically takes advantage of specific software libraries~\footnote{\label{ibm_open}\url{https://github.com/open-power}} to implement the real-time execution environment required to run power management policies under tight timing constraints.
Many-core power management demands fine-grained control of the operating points of the \glspl{pe}~\cite{PERCOREPSTATE} to meet a given processor power budget while minimizing performance penalties. The control policy has to provide fast and predictable responses to promptly handle the incoming requests from the \gls{os} or \gls{bmc} and prevent thermal hazards.
A flexible and scalable way to sustain these computationally intensive operations is required to provide accurate control per core and to support more advanced control policies, such as those based on model-predictive control~\cite{BARTOLINI_MPC}.

Furthermore, simultaneous \gls{hw} and \gls{sw} development of \gls{pcs} and \gls{pcf} has to be coupled with a dedicated co-design and validation framework within the \textit{controlled system}. Indeed \gls{pcs} and \gls{pcf} performance directly depends on their interaction with the physical state of the controlled processors --- temperature, power, workload, and control decisions from the \glspl{hlc}. 
Such a physical state time-scale model requires near real-time speed to be meaningful.
This involves, on the one hand, the design of adequate on-chip interfaces between the controller to be validated --- \gls{isut} --- and the surrounding system. 
On the other, a proper \textit{virtual} representation of the system surrounding the \gls{isut} that models physical components on a target computer~\cite{HIL_ISUT}, called \textit{plant model}. 
In a design relying on dedicated \glspl{pcs}, the controlled plant is a complex \gls{mimo} \gls{mpsoc} with \glspl{ap} that is often not available during the design phase of the \gls{pcs} to be integrated, thus being replaced by a thermal and power model encapsulating floorplan, power and thermal information of the \gls{cpu} under a particular application workload.

Today's advanced \gls{hesoc} integrate both single/multiprocessors \glspl{mcu} and configurable hardware components on the same die (\gls{fpga}-\gls{soc}). 
Their native integrated and configurable structure makes them ideal options for closed-loop emulation of on-chip \glspl{pcs} targeting advanced power management.

Therefore, this paper's central idea is to fulfill a twofold need: on the one hand, the design of a capable \gls{pcs} architecture optimized for handling a per-core, reactive thermal and power control strategy within the required power budget and timing deadlines. 
On the other, the design of an agile, closed-loop framework for the joint \gls{hw}/\gls{sw} development of integrated control systems, 
leveraging the capabilities of modern \gls{hesoc} platforms.

To the best of the authors' knowledge, this work proposes the first research platform where the full-stack (\gls{hw}, \gls{rtos}, \gls{pcf} and power/thermal emulation setup) required for \gls{llc}-driven on-chip power and thermal management is released open-source~\footnote{\url{https://github.com/pulp-platform/control-pulp}}.
\revision{Unlike traditional Linux-capable \gls{smp} multi-core systems, as a research platform for on-chip power and thermal management, ControlPULP aims to keep the design's \gls{hw} and \gls{sw} complexity to a 32-bit manager and 32-bit \gls{pmca} with \gls{rtos} support. This design choice better fits the resource constraints of a controller embedded in the \textit{uncore} domain of a larger \gls{cpu} that is assumed to integrate out-of-order, massive application-class processing elements already.}

The manuscript significantly extends the conference work in~\cite{CPULP-SAMOS} by detailing and jointly validating the power/thermal model on the \gls{hesoc}-\gls{fpga} with the \gls{pcs} and \gls{pcf}.
Overall, the work provides the following contributions: 

\begin{enumerate}
    \item We design an end-to-end RISC-V parallel \gls{pcs} architecture named ControlPULP, based on open RISC-V cores and hardware IPs~\cite{PULP}. ControlPULP is the first fully open-source \gls{pcs} with a configurable number of cores and hardware resources to track the computational requirements of the increasingly complex power management policies of current and future high-performance processors (Sec.~\ref{sec:architecture}), as well as the first proposed in the RISC-V community.
    ControlPULP integrates a \textit{manager core} coupled with a multi-core \gls{pmca} (cluster) with per-core FPUs for reactive control policy step computation.
    The cluster subsystem hosts a \gls{dma} engine to accelerate the data transfers from on-chip sensors. This enables data acquisition with 2D stride patterns, a crucial capability when reading from \gls{pvt} sensors with equally spaced address mapping (Sec.~\ref{subsec:dma-test}).
    \item We tailor ControlPULP to meet real-time power management requirements. 
    The architecture integrates a fast Core Local Interrupt Controller (RISC-V CLIC~\cite{CLIC}) tasked to process the interrupt messages associated with \gls{os}- and \gls{bmc}- driven commands, as well as a low latency predictable interconnect infrastructure (Sec.~\ref{subsec:scmi-test}).
    \item We demonstrate the end-to-end capabilities of ControlPULP with a case study on the control algorithm \gls{qos}. 
    We develop a closed-loop evaluation framework based on the concept of \glspl{hesoc} for real-time characterization of the control policy while simultaneously fast-prototyping the underlying hardware (Sec. \ref{sec:fpga-map}). The framework relies on a \textit{power, thermal and performance model} of the controlled \gls{cpu} as \textit{plant}, which is paired with workload instruction traces and the \gls{pcf} to realize the closed-loop.
    This approach enables a multi-step validation and characterization environment ranging from \gls{sil} and \gls{mil} to \gls{hil} abstractions (Sec. \ref{subsec:pcf-control-eval}).
    \item Finally, we benchmark the reactive control algorithm against one of the few freely-accessible industry-graded \gls{sota} control policy, IBM's OpenPower, in a \gls{mil} simulation, achieving 5\% more precise set-point tracking (Sec.~\ref{subsec:pcf-control-eval}).
\end{enumerate}

\section{Related Work}\label{sec:related_work}

There is little publicly available information on \glspl{pcs} architectures and their \gls{hw}/\gls{sw} interface. 
Table \ref{tab:pcs_sota} summarizes the leading solutions from academia and industry. 

\begin{table*}[ht]
    \caption{Comparison among existing proprietary and freely-available \gls{pcs} from industry and academia.}
    \begin{center}
    \renewcommand{\arraystretch}{1.3} 
    \resizebox{\linewidth}{!}{
    \begin{tabular}{cc|cccc}

        \textbf{\gls{pcs}} & \textbf{Provider} & \textbf{HW} & \textbf{\thead{\gls{pcf} \\ scheduling}} & \textbf{\thead{Programmable \\ accelerator}} & \textbf{\thead{Openness \\ (HW/\textcolor{blue}{SW})}} \\
        
        \hline 
        \multicolumn{6}{c}{\textbf{Industry}} \\ 
        \hline 
        
        PCU  & Intel & 1-core, HW FSMs & n.a. & \xmark & \xmark \hspace{0.3pt} \textcolor{blue}{\xmark} \\
        SCP, MCP  & Arm & 2-cores & SW FSMs & \xmark & \xmark \hspace{0.3pt} \textcolor{blue}{\cmark}  \\
        SMU & AMD & 1-core & n.a. & \xmark & \xmark \hspace{0.3pt} \textcolor{blue}{\xmark} \\
        \textbf{OCC} & IBM & 1-core, microcode engines & SW FSMs & \xmark & \xmark \hspace{0.3pt} \textcolor{blue}{\cmark} \\

        \hline 
        \multicolumn{6}{c}{\textbf{Academia}} \\ 
        \hline 
        
        Bambini et. al~\cite{gap8-hil} & academic & 1-core & RTOS & \xmark & \xmark \hspace{0.3pt} \textcolor{blue}{\cmark} \\
        \textbf{This work} & academic & 1-core, cluster accelerator & RTOS & \cmark & \cmark \hspace{0.3pt} \textcolor{blue}{\cmark} \\

        \hline
        \end{tabular}
    }
    \label{tab:pcs_sota}
    \end{center}
\end{table*}

Intel's \gls{pcu}, introduced with Nehalem microprocessor~\cite{NEHALEM}, is a combination of dedicated hardware state machines and an integrated microcontroller~\cite{SANDYBRIDGE}. It provides power management functionalities such as \gls{dvfs} through voltage-frequency control states (P-states and C-states), selected by the HW (\textit{Hardware-Managed P-States}). The \gls{pcu} communicates with the \glspl{pe} with a specialized link through \textit{power management agents}. Intel's main control loop runs at $500 \mu s$~\cite{schone_intel}. 

AMD adopts a multiple power controller design, with one \gls{smu} for each \gls{cpu} tile (group of cores) in a Zeppelin module. All \glspl{smu} act as slave components, monitoring local conditions and capturing data. One of the \glspl{smu} also acts as a master, gathering all information from the slave components and then choosing the operating point for each core~\cite{AMD}.

Arm implements two independent \glspl{pcs} based on the Arm Cortex-M7 microcontroller, \gls{scp} and \gls{mcp}. The \gls{scp} provides power management functionality, while the \gls{mcp} supports communications functionality. 
In Arm-based \glspl{soc} the interaction with the \gls{os} is handled by the \gls{scmi} protocol~\cite{SCMI}. \gls{scmi} provides a set of \gls{os}-agnostic standard \gls{sw} and \gls{hw} interfaces for power domain, voltage, clock, and sensor management through a shared, interrupt-driven mailbox system with the \gls{pcs}.

The IBM \gls{occ}, introduced with Power8 microprocessor, is composed of 5 units: a central PowerPC 405 processor with 768 KiB of dedicated SRAM and four microcode general-purpose engines (GPEs). The latter are responsible for data collection from \gls{pvt} sensors, performance state control and \gls{cpu} stop functions control (\textit{PGPE} and \textit{SGPE}) respectively. IBM \gls{occ}'s \gls{pcf} is called OpenPOWER, and has a periodicity of $250 \mu s$~\cite{ibm_occ}. It relies on a \textit{frequency voting box} mechanism to select a frequency for each core conservatively based on the minimum input - highest \textit{Pstate} - from several independent power-control (\textit{control vote}) and temperature-capping (\textit{thermal control vote}) features.
The \textit{thermal control vote} consists of one \gls{pid} that reduces the frequency of each core based on the temperature of the hottest \gls{pe}.
Furthermore, similarly to Arm's \gls{scmi} standard, IBM's \gls{occ} relies on a \textit{command write attention/interrupt mechanism} to notify the \textit{PGPE} of an incoming asynchronous command/request to be processed\footnotemark[\getrefnumber{ibm_open}], such as the desired \textit{PState}. \textit{PGPE} arbitrates this information with the voting box output from the PowerPC 405 according to a minimum \textit{PState} policy.

Last, Bambini et al.~\cite{gap8-hil} show that a single-core, RISC-V-based microcontroller can execute similar reactive control algorithms with a control loop of 500$\mu$s. The work relies on the SPI interface to conduct the main periodic task and lacks support for essential \gls{hw} \glspl{ip} such as \gls{dma} and \gls{fpu}.

All the \gls{sota} power controllers lack the flexibility and scalability of a multi-core architecture supported by adequate I/O bandwidth from/to on-die and off-die power monitors and actuators coupled with fast interrupt handling hardware for \glspl{hlc} (\gls{os} and \gls{bmc}) commands dispatching, which is the critical innovation provided by ControlPULP.

It is essential to notice that \glspl{pcs} design is only half of the coin, as its performance depends on the \gls{pcf} and real-time performance achieved in closed-loop. Several works have targeted the emulation of 'in-field' power management algorithms, but none of them has validated the \gls{pcs} and \gls{pcf} designs jointly (co-designed).

Atienza et al.~\cite{ATIENZA_FPGA, ATIENZA_FPGA_2} propose a thermal emulation framework where a generic \gls{mpsoc} is implemented on \gls{fpga}. A host computer executes the \textit{thermal model}, which is driven by real-time statistics from processing cores, memories, and interconnection systems emulated on the \gls{fpga}. With the increasing number of cores in modern \glspl{cpu}, an \gls{fpga} approach that implements the entire \gls{mpsoc} is not feasible and incurs resource partition with high platform costs.
Beneventi et al.~\cite{BENEVENTI_ONLINE} design a similar closed-loop approach where a subset of the Intel Single-Chip-Cloud computer's \glspl{pe} execute the \textit{thermal model} while receiving online workload information from the remaining \glspl{pe}. While employing actual and commercial hardware in the emulation, the work focuses on the \gls{hlc} only and with a software-centric methodology, being \gls{hw}/\gls{sw} co-design of the \gls{llc} prevented by the inaccessibility of the underlying \gls{hw}.
 
Beyond these approaches, our work aims at co-designing the \gls{llc} (\gls{pcs}) and \gls{hw}/\gls{sw} components of the controlled plant to assess "in-field" performance.
This step is well understood in the design flow of classic control systems targeting automotive, avionics, and robotics domains, where progressively more realistic simulations of the plant in the closed-loop are coupled with the introduction of the actual hardware controller (\gls{hil}) and checked against model-based closed-loops (\gls{sil}, \gls{mil}) from early development stages of the control design.

A taxonomy of the various design possibilities for a control system that adapts well to the scenario of this work is provided in~\cite{HIL_AUTOMATION}.
On the controller side, an \gls{fpga} emulation approach provides the benefit of testing the control firmware developed during \gls{sil}/\gls{mil} on the actual hardware controller with the guarantee of cycle-accurate simulation. The latter is required to achieve fine-grained hardware observability and controllability~\cite{HIL_ISUT} and one-to-one correspondence between the \gls{rtl} source and its \gls{fpga} mapping in terms of clock cycles (Direct \gls{fame} systems~\cite{FAME}).
On the plant side, a virtual simulation of the plant is preferred to an \gls{fpga}-based approach, which incurs high development costs due to the complexity of the whole plant system to be emulated. For this purpose, \glspl{mcu} are the solution adopted by the industry nowadays: they are cheap, general-purpose and with a standardized and automated software development process.

The combination of \glspl{mcu} and \glspl{fpga} flexibility takes the best of both worlds and is supported by modern, \glspl{cots} \gls{fpga}-\glspl{soc} platforms such as Xilinx Zynq Ultrascale+ and Versal families, making it the solution of choice adopted in our work. 

\section{The ControlPULP platform}\label{sec:architecture}

ControlPULP extends commercial controllers' single-core architecture, providing the first multi-core RISC-V \gls{pcs} architecture.
To make the paper self-contained, in the following, we first provide a high-level overview of a generic \gls{hpc} \gls{cpu} that integrates the \gls{pcs}. We detail the platform's software stack, which helps understand the control policy flow, the interaction between the controlled plant and the controller, and the terminology adopted throughout the manuscript (Sec. \ref{subsec:fw-archi}).
Finally, we detail ControlPULP's hardware architecture and design trade-offs (Sec. \ref{subsec:cpulp-archi}) to implement the control algorithm.

\subsection{Controlled plant and Power Control Firmware}\label{subsec:fw-archi}

\begin{figure*}[htb]
    \centering
    \includegraphics[width=\linewidth]{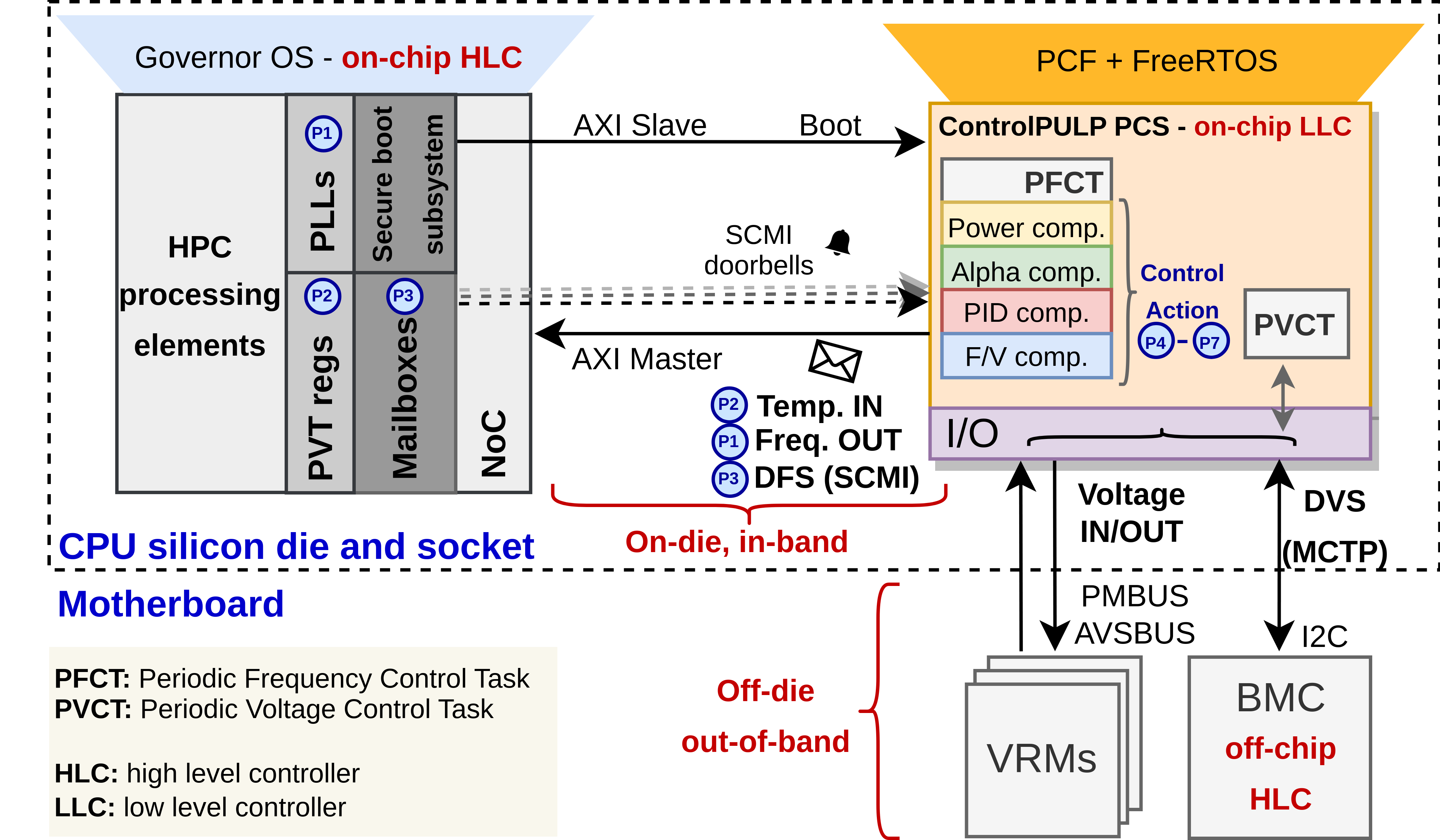}
    \caption{High-level overview of the system. We highlight on-chip and off-chip \glspl{hlc} (\gls{os} and \gls{bmc}), the \gls{llc} (ControlPULP) and the \gls{mimo} IO interfaces. 
    Furthermore, the figure details the \gls{pcf} phases described in Sec. \ref{subsec:fw-archi}}
    \label{fig:fw-sketch}
\end{figure*}

Fig. \ref{fig:fw-sketch} depicts the high-level structure of the \gls{hpc} \gls{cpu} silicon die. From a control perspective, we distinguish between the controlled system (\textit{plant}) and the on-chip \gls{llc} controller, namely ControlPULP. 
Furthermore, the figure illustrates the environment surrounding the \gls{cpu} socket hosting the silicon die, namely the motherboard, with off-chip actors, such as \glspl{vrm} and \gls{bmc}. The \gls{os} running on the \glspl{pe}, as well as the off-chip \gls{bmc}, are the two \glspl{hlc}.

We assume the controlled plant is a many-core \gls{hpc} \gls{cpu} with 72 application-class \glspl{pe}, and exposes hardware mailboxes through a \gls{noc} interconnect system.
While the mailboxes mediate the \gls{dfs} commands (e.g. target frequency) dispatched on-chip by the governor OS \gls{hlc} to the \gls{llc}, power management protocols such as \gls{pmbus}, \gls{avsbus} and \gls{mctp} mediate \gls{dvs} commands (e.g. power budget threshold) from the \gls{bmc} \gls{hlc}, as detailed in Sec. \ref{subsubsec:io-interf}.

To clarify the terminology in Fig. \ref{fig:fw-sketch}, \textit{on-die} designates any element of the \gls{hpc} \gls{cpu} that resides on the chip die, such as \gls{pvt} sensors and registers, frequency actuators, and mailboxes. \textit{In-band} services refer to \gls{scmi}-based interaction and \gls{pvt} data acquisition. Lastly, \textit{off-die} indicates \glspl{vrm} communication and \gls{bmc} requests through \textit{out-band} services.

\begin{figure*}[htb]
    \centering
    \includegraphics[width=0.4\linewidth]{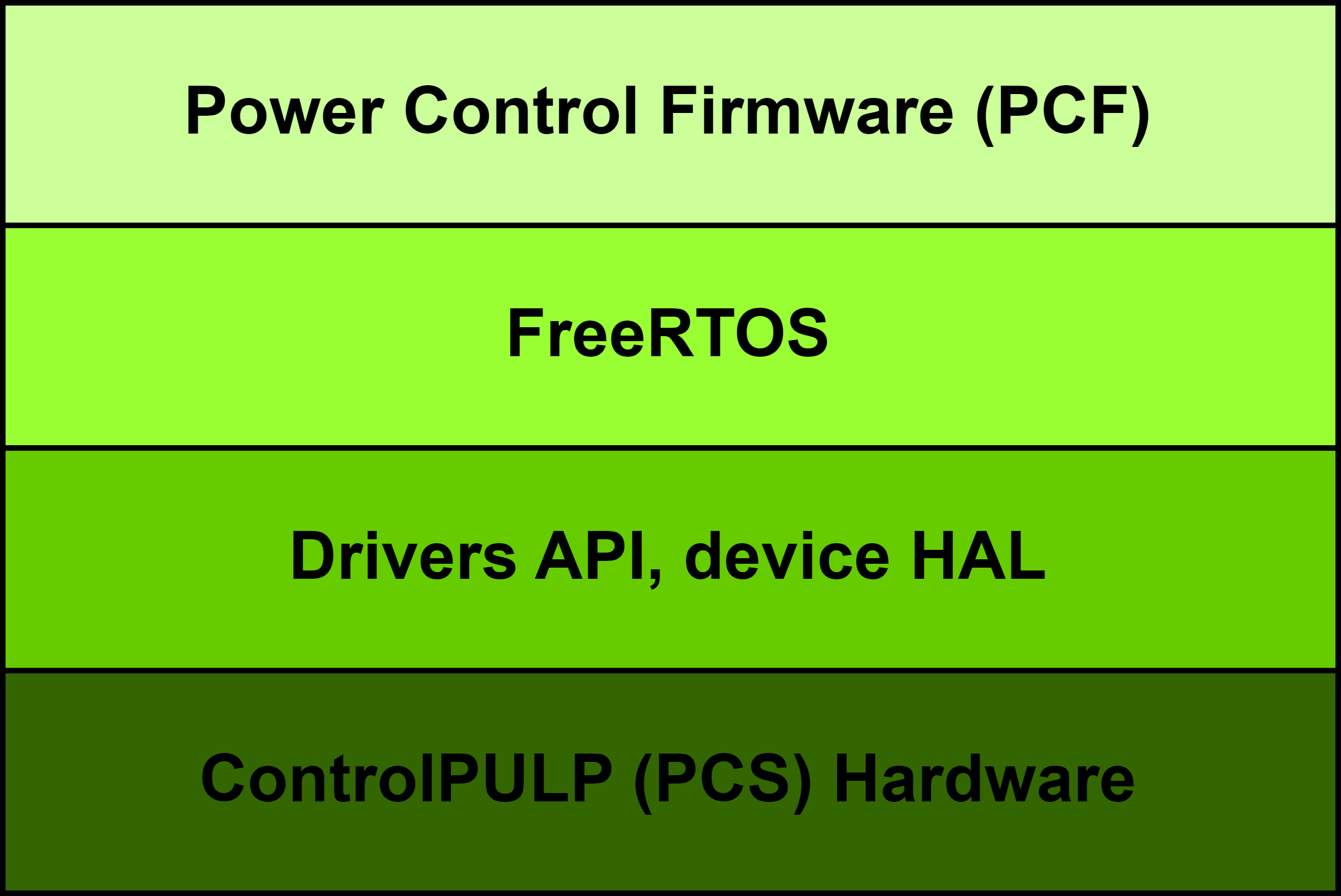}
    \caption{ControlPULP software stack. The application control policy (\gls{pcf}) executes on top of FreeRTOS, which controls the hardware with target-specific drivers and \gls{hal} \glspl{api}}
    \label{fig:cpulp_sw_stack}
\end{figure*}

The \gls{pcf} executes the thermal and power control functions and manages on-die and off-die communications and data transfers. Real-time priority-driven scheduling with static task priorities and preemption is required to manage the control functions. In this work, we use FreeRTOS, an industry-grade, lightweight, and open-source operating system for microcontrollers. The software stack of the proposed platform is shown in Fig. \ref{fig:cpulp_sw_stack}.

The \gls{pcf} routine implements a two-layer control strategy~\cite{gap8-hil}, managed by the FreeRTOS scheduler as two periodic tasks characterized by multiple harmonic frequencies: the \textbf{\gls{pfct}} --- 2 kHz, i.e. 500$\mu s$ --- and the \textbf{\gls{pvct}} --- 8 kHz, i.e. 125$\mu s$. 
Splitting the control routine into two tasks grants more fine-grained control actions and helps meet different performance requirements and sensors-update frequencies. For instance, power changes rapidly due to instruction-level variation of the effective switching capacitance of the computing units, while temperature variations are slower. The control policy has to handle these widely split time scales. Furthermore, \glspl{vrm} generally update more frequently than temperature sensors. 

The \gls{pfct} is the main control task. It receives the desired clock frequency operating point for each processing element from the \gls{os} \gls{hlc} governor as well as a power budget threshold from the \gls{bmc} \gls{hlc} via the \gls{pvct}, and computes the optimal frequency/voltage pair to meet the physical and imposed constraints of the system. The task is then responsible for dispatching the controlled frequency by directly accessing the \gls{cpu} PLLs, as from Fig. \ref{fig:fw-sketch}.
The \gls{pfct} executes a two-layer control strategy~\cite{gap8-hil} consisting of a \textit{power dispatching layer} and a \textit{thermal regulator layer}.
\gls{pfct}'s control step $n$ comprises several phases, illustrated in Fig. \ref{fig:fw-sketch}: \textbf{(P1)} allocate the controlled clock frequency computed at step $n-1$ to each core; \textbf{(P2)} read the \gls{pvt} sensor's registers and the workload characteristics from each core; \textbf{(P3)} obtain commands and information on the constraints (\gls{dvfs} operating points, power budget) from the \gls{os} and the \gls{bmc}; \textbf{(P4)} compute the estimated power for each core and the total consumed power of the system; \textbf{(P5)} apply a power capping algorithm, such as \textit{alpha}~\cite{gap8-hil} when the total power exceeds the power budget constraint; \textbf{(P6)} further reduce the power of each core through \glspl{pid} computation when the temperature at phase (P2) exceeds the threshold; \textbf{(P7)} compute both the frequency and voltage to be applied to the controlled system. Throughout this manuscript, we name \textit{control action} the computational body of the \gls{pcf} execution (P4)-(P7). 

The \gls{pcf} does not provide per-core voltage control but groups \glspl{pe} in coarse-grained voltage domains. The \gls{pvct} is responsible for detecting the changes in the system's power consumption. It periodically reads the power consumption of the voltage rails --- associated with each voltage domain --- from the \glspl{vrm} and programs micro-architectural power/instruction throughput capping interfaces (if supported by the \glspl{pe}). Lastly, it modifies the \gls{pfct}'s power budget threshold as requested by the \gls{bmc}. 
Even though the \gls{pfct} computes both optimal frequency and voltage, it only dispatches the frequencies to apply at phase (P1) in step $n+1$. In contrast, the \gls{pvct} dispatches the voltages at step $n$ (one iteration before), hence the names chosen for the two tasks.

\subsection{System architecture}\label{subsec:cpulp-archi}

\begin{figure*}[t]
    \centering
    \includegraphics[width=\linewidth]{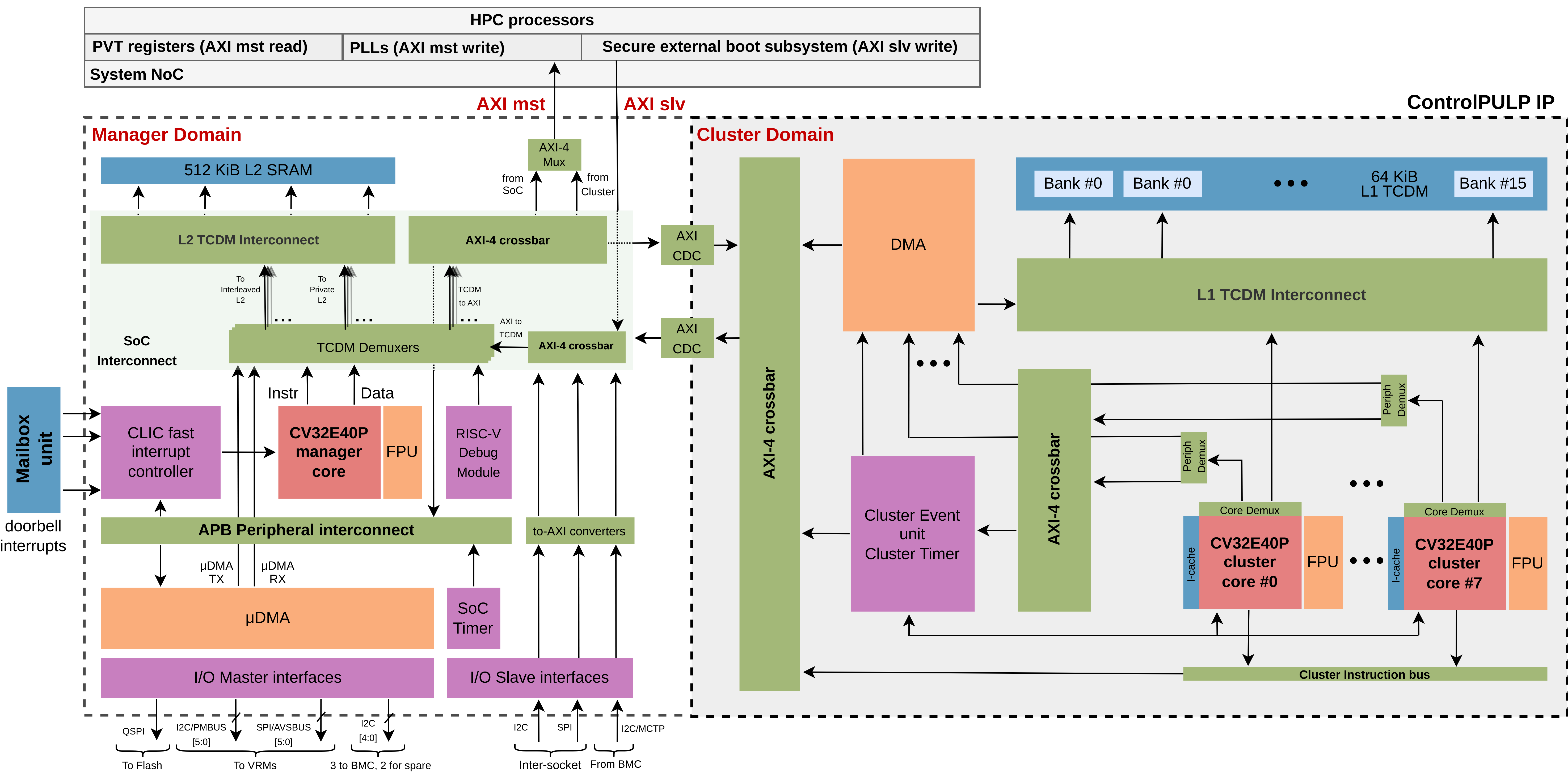}
    \caption{ControlPULP hardware architecture. On the left, the \textit{manager domain} with the \textit{manager core} and surrounding peripherals. On the right, the \textit{cluster domain} accelerator with the eight cores (workers)}
    \label{fig:cpulp_archi}
\end{figure*}

Fig. \ref{fig:cpulp_archi} provides a block diagram overview of ControlPULP. 
The top-level subsystem of the design is the \textit{manager domain}, consisting of a CV32E40P open-source~\footnote{\url{https://github.com/openhwgroup/cv32e40p}} industry-grade processor core and a set of system I/O interfaces to interact with external peripherals and memory-mapped devices (Sec. \ref{subsubsec:io-interf}). The primary micro-controller-like subsystem is also a recurrent element in the \gls{sota} designs surveyed in Sec. \ref{sec:related_work}.

\subsubsection{Real-time and predictability}
In the following, we discuss the architectural design decisions concerning RAM banking and interrupt processing taken to make the design more suitable for real-time workloads.

\paragraph{L2 memory banks constant access time}\label{par:l2_mem_bank}
The L2 RAM, which is the RAM block connected to the \textit{manager domain} and system I/O interfaces, is sized to 512 KiB, enough to fit the whole firmware binary and data so that no swapping is required. 
The L2 RAM comprises six banks. The access time to each bank is constant when there are no access conflicts. 
Two of these banks are marked private to prevent \gls{dma} transfers from peripherals and other components from disturbing the manager core's instruction and data fetching. The manager core has exclusive access to those.

\paragraph{Low constant latency core-local (CLIC) interrupt controller}\label{par:clic}
Provided the need for a real-time and predictable architecture, we extend the original \gls{clint}, compliant with RISC-V privileged specifications~\cite{RISCVSPEC2}, with the newly proposed \gls{clic}, currently under ratification by the RISC-V community.
We employ an open-source implementation of the \gls{clic}~\footnote{\url{https://github.com/pulp-platform/clic}}~\cite{balas2023cv32rt} that reflects the latest status of the RISC-V \gls{clic} draft specifications~\cite{CLIC}. The integration process includes the addition of CSRs registers in the processor's micro-architecture as per specifications.

The \gls{clic} introduces several improvements to the standard \gls{clint} to achieve faster interrupt handling. Among those are dedicated memory-mapped registers for software configurable interrupt priority and levels at the granularity of each interrupt line, runtime-configurable interrupt mode and trigger type, and support for interrupt preemption within the same privilege level (interrupt nesting).
\Gls{shv} enables the programmer to optimize each incoming interrupt for either faster response (\textit{vectored} mode, when each interrupt service routine has a reserved entry in the interrupt table) or smaller code size (\textit{direct} mode, when each interrupt traps to the same exception handler address).
Lastly, with the \gls{clic}, local interrupts can be extended to 4096 lines instead of being limited to the processor's XLEN (32-bit for CV32E40P). 

In this work, we implement 256 local interrupt lines coming to ControlPULP from the mailbox infrastructure~(Sec. \ref{subsubsec:io-interf}).
The \gls{clic} configuration helps reduce the interrupt response latency and is capable of entering the interrupt handler within 30 clock cycles (Sec.~\ref{sec:results}). This is a crucial property to increase responsiveness on external, agent-driven requests.

\subsubsection{Cluster accelerator}
To meet the computational demands of the control algorithms, in particular, when scaling to a large number of controlled high-performance \glspl{pe} and improving the control performance, we opt for a flexible programmable accelerator, namely a cluster of RISC-V CV32E40P cores --- referred to as \textit{workers} in this manuscript --- tightly coupled to 64 KiB RAM (L1) and a \gls{dma} engine. The accelerator is represented in Fig. \ref{fig:cpulp_archi} as \textit{cluster domain}.

\paragraph{Multi-core computing system}\label{par:cluster_multi}
Control algorithms (Sec.~\ref{subsec:fw-archi}) can be parallelized on the cluster domain (Sec.~\ref{subsec:fw-test}). This guarantees a high grade of flexibility on the software development side, and is in sharp contrast with hardwired control logic featured in \gls{sota} controllers (Sec.~\ref{sec:related_work}), which lack flexibility.
The \textit{manager core} offloads the control algorithm to the team of workers in the cluster. Each worker features a private instruction for the instructions stored in the main L2 memory and accesses L1 through a single-cycle latency logarithmic interconnect.

In the most straightforward parallelization scheme, a worker computes the control action (Sec. ~\ref{subsec:fw-archi}) for a subset of the controlled cores. 
The number of workers in the cluster is parametric. In the following, we consider eight cores to demonstrate scalability. Each core in the cluster features an \gls{fpu} with a configurable number of pipeline stages. In our instantiation, we use one internal pipeline stage, which is sufficient to meet our frequency target (Sec. \ref{subsec:cpulp-synth}). Furthermore, Montagna et al.~\cite{LOWPOWERTRANS} show that this configuration achieves high performance and reasonable area/energy efficiency on many benchmarks.

\paragraph{2-D DMA transfer engine}\label{par:cluster_dma}
The \textit{cluster domain} integrates a multi-channel \gls{dma} with direct access to L1 RAM and low-programming latency ($62$ clock cycles, Sec. \ref{subsec:dma-test}).
The \gls{dma}'s main task is to provide direct communication between L2 and L1 memories in parallel and without intervention from the \textit{manager} or \textit{cluster} domains~\cite{DMA}.

We tailor the \gls{dma}'s capabilities to suit the control policy use case by (i) directly routing the cluster \gls{dma} to the \gls{pvt} sensors registers through the outgoing AXI master interface, which guarantees flexibility by decoupling data transfers and computation phases, (ii) exploiting 2-D transfers for equally spaced PVT registers accesses and (iii) increasing the number of outstanding transactions (up to 128) to hide the latency of regular transfers.

Commercial \glspl{pcs} described in Sec. \ref{sec:related_work} also decouple the actual computation from data acquisition. For instance, according to available \gls{sota} information, IBM's \gls{occ} employs general-purpose cores (GPEs) tasked to read \glspl{pe}'s data and temperatures instead of a dedicated data mover engine with reduced programming interface overhead~\cite{ibm_occ}.

\subsubsection{System I/O interfaces}\label{subsubsec:io-interf}

\paragraph{AXI4 interfaces}
ControlPULP features two AXI4 ports, one \textit{master} and one \textit{slave}, with 64-bit W/R, 32-bit AW/AR wide channels. They play a crucial role in the design and guarantee low-latency communication with the controlled system.
The AXI slave drives the booting process of the \gls{pcs}. In the high-level structure depicted in Fig. \ref{fig:fw-sketch}, an external, secure subsystem is responsible for loading the \gls{pcf} binary into ControlPULP's L2 SRAM by mapping to a region of the L2 SRAM.
The AXI master is the transport layer over which the \gls{pcs} collects \gls{pvt} sensors data and power policy target requirements. On the other hand, it dispatches the computed optimal operating point to the \glspl{pe} during the control policy (Sec. \ref{subsec:fw-archi}). Its address space is reachable from both the \textit{manager} and \textit{cluster} domains.

\paragraph{Mailbox-based SCMI communication}

ControlPULP adopts and implements the Arm standard \gls{scmi} protocol to handle external power, performance, and system management requests from the \gls{os} \gls{hlc}. 
\gls{scmi} allows an \gls{os} kernel that supports \gls{scmi} to interact with ControlPULP without requiring a bespoke driver. Furthermore, the design of the \gls{scmi} protocol reflects the industry trend of delegating power and performance to a dedicated subsystem~\cite{SCMIWHITE}.
\gls{scmi} involves an interface channel for secure and non-secure communication between a caller (named \textit{agent}, i.e., one processing element of an \gls{hpc} \gls{cpu}) and a callee (named \textit{platform}, i.e., ControlPULP). The latter interprets the messages delivered by the former in a shared memory area (mailbox region, Fig. \ref{fig:fw-sketch}) and responds according to a specific protocol.
The proposed \gls{pcs} implements an interrupt-driven transport mechanism through the \gls{clic}. In our use case with 72 controlled cores, the platform can process up to 144 secure, and non-secure interrupt notifications.

We design hardware mailboxes as the transport layers for the \gls{scmi} communication mechanism. The shared memory region is implemented according to the single-channel memory layout described in the specifications.
We reserve a space of 8B for the implementation-dependent payload field, totaling 40B per channel.
Each channel is associated with one outgoing interrupt line in the agent-platform direction (doorbell). 
The agent identifier is encoded with the message such that more agents can use the same channel as the notification vehicle. 
Hence, the number of interrupts dispatched by the mailbox system can be smaller than the number of agents, with the benefit of reducing the area of the interrupt controller's configuration register file. The agent identifier is not described in the official specifications and is mapped to a reserved field of the single-channel memory layout.
With 64 channels as employed in this work, the shared memory region has a size of about 2kiB.

\paragraph{I/O peripherals for voltage rails power management}
ControlPULP integrates a peripheral subsystem in the \textit{manager domain}, where an I/O data engine unit (named $\mu$DMA \gls{ip}) enables autonomous communication between off-die elements and the L2 SRAM with low programming overhead for the manager core.
In this work, we upgrade the peripheral subsystem with industry-standard power management interfaces to handle off-die communication services.
The \gls{pcs} integrates 6 \gls{avsbus} and \gls{pmbus} interfaces towards \glspl{vrm}. The \gls{pmbus} and \gls{avsbus} bus protocols extend I2C and SPI to monitor voltage and power rails digitally, preserving optimal speed/power consumption trade-off.
5 I2C master/slave interfaces manage the communication with the \gls{bmc} and other board-level controllers. The slave interfaces transfer \gls{dvs} operating points (power budget) from the \gls{bmc} according to the \gls{pldm} and \gls{mctp} transport layer protocols. 

\section{FPGA-based HIL thermal, power, performance and monitoring emulation framework}\label{sec:fpga-map}

This section provides an end-to-end description of the \gls{hesoc} based \gls{hil} emulation methodology.
The closed-loop approach that we implement introduces the actual \gls{llc} executing the \gls{pcf} as \gls{isut} and relies on a \textit{thermal, power and performance model} for the controlled \gls{cpu} plant. 
While the control literature refers to such a setup as \gls{pil} or \gls{fil}, others already define a closed-loop as \gls{hil} when the connection between the integrated system under test and the plant reflects the actual hardware interface of the final manufactured silicon, without relying on a virtual representation~\cite{HIL_ISUT}. Since this is the case for the present work, we adopt the \gls{hil} terminology.

\subsection{HIL system and PCS mapping on FPGA}\label{subsubsec:hil-isut}

\begin{figure*}[t]
    \centering
    \includegraphics[width=\linewidth]{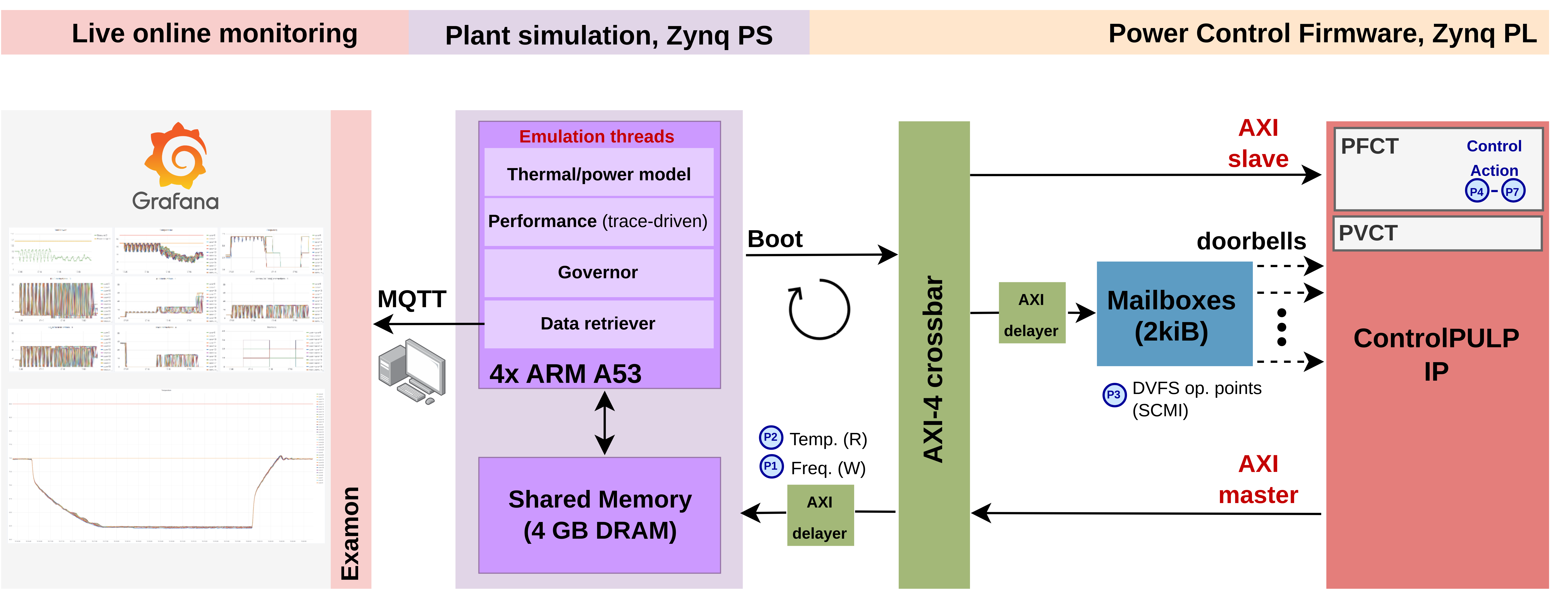}
    \caption{\gls{hil} test procedure on an \gls{fpga}-\gls{soc} with ControlPULP}
    \label{fig:hil-concept}
\end{figure*}

The Xilinx Ultrascale+ \gls{fpga} family is widely adopted in the heterogeneous computing domain~\cite{HEROv2}. It features a \gls{ps}, or host computer, and a \gls{pl}, namely the configurable \gls{fpga}, integrated on the same physical die.
Fig. \ref{fig:hil-concept} provides an overview of the \gls{fpga} based emulation framework designed in this work and its main actors.

The \gls{ps} consists of an industry-standard, application class, quad-core, 64-bit Armv8\textsuperscript{\textregistered} Cortex-A53 \textit{Application Processing Unit} with 32 KiB L1 instruction and data cache per core and a 1 MiB L2 cache shared by all four cores --- \gls{ocm} ---, clocked at 1.2 GHz, a dual-core Cortex-R5F \textit{Real Time Unit}, and Mali\textsuperscript{\texttrademark}-400 MP2 GPU based on Xilinx's 16nm FinFET.

We implement ControlPULP on the ZCU102 \gls{pl} with Xilinx Vivado 2022.1. The \gls{ps} interfaces with ControlPULP through the AXI4 master and slave ports (Sec. \ref{sec:architecture}) and provides it with an external reset and 125 MHz clock, which is internally converted to 50 MHz system clock by the \gls{isut}.
Table \ref{tab:fpga_resources} shows the board's resource utilization inferred by the implementation. Overall, ControlPULP fits the available resources with almost 97\% utilization. 

\begin{table}[htb]
    \caption{ControlPULP resource utilization on the Xilinx UltraScale+ ZCU102.}
    \begin{center}
    \renewcommand{\arraystretch}{1.2} 
    \begin{tabular}{cccc}
        \hline
        \textbf{Resource} & \textbf{Utilization} & \textbf{Available} & \textbf{Utilization [\%]} \\
        \hline
        \textbf{LUT}  & 265718 & 274080 & \textbf{96.95}\\
        \textbf{FF}   & 153155 & 548160 & 27.94\\
        \textbf{BRAM}   & 278 & 912 & 30.48\\
        \textbf{DSP}   & 93 & 2520 & 3.69\\
        \textbf{I/O}  & 83 & 328 & 25.30\\
        \hline
        \end{tabular}
    \label{tab:fpga_resources}
    \end{center}
\end{table}

The communication between the plant simulation and the \gls{isut} is regulated through a 4 GiB off-chip DDR4 DRAM provided on the board, which allows them to exchange data and realize the feedback loop with up to 19.2 GB/s of bandwidth.

\subsection{Plant simulation}\label{subsubsec:hil-plant_sim}

In the considered control scenario, the plant provides a thermal, power, performance, and monitoring framework capable of simulating the power consumption and temperature of a high-end \gls{cpu} processor.
The plant simulation is programmed in \textit{embedded C}. Its functionalities are split into several threads employing \texttt{pthread} Linux libraries to exploit multi-core parallelism. In the following, we refer to Fig. \ref{fig:hil-concept} and detail the threads' organization and interaction with the \gls{isut}.

The \textbf{\textit{thermal and power model thread}} is the leading simulation thread. It carries out the computation of the thermal and power model with a periodicity of $1 \mu s$ which, according to the literature~\cite{Beneventi, CCTA}, is fast enough to capture the thermal dynamics, i.e., three orders of magnitude faster than the fastest thermal time constant $\sim1ms$, and to simulate power spikes and oscillations that are not filtered by the hardware power delivery network (up to $\sim100 \mu s$~\cite{PDN}).
The thread takes as input the values controlled and dispatched by the \gls{pcs} controller (frequency and voltage) and the modeled workload from the \textit{performance model thread} to compute the average consumed power and the temperature of each core.
The average consumed power is computed with an algebraic model of the cores, which includes core-to-core variability and noise, according to~\cite{CCTA}:

\begin{equation}\label{eq:power}
    P_{i} = P_{i,\mathrm{static}} + P_{i, \mathrm{dynamic}} = \kappa(T_{\mathrm{Si},i}) \cdot (I_{\mathrm{cc},i} \cdot V_{\mathrm{dd}} + (C_{\mathrm{eff},i} \cdot f_{i} \cdot V_{\mathrm{dd}}^{2}))
\end{equation}

\noindent where $\kappa$ represents the temperature dependency of the computed power, and $C_{\mathrm{eff}}$ is the equivalent effective capacitance of the controlled \gls{cpu}.
The temperature is simulated through a discrete state space model, which considers the temperature and instantaneous power of the neighboring simulated cores. 
Coefficients are extracted from a commercial multi-node RISC-V cluster capable of providing an \gls{hpc} production stack, Monte Cimone~\cite{MONTECIMONE}.

The \textbf{\textit{performance model thread}} assists the \textit{thermal and power model thread} by providing the workload characteristics for the next interval of time.
We rely on a \textit{trace-driven} approach for the simulated workload for practical reasons: (i) the traces could be extracted from accurate benchmarks such as PARSEC~\cite{PARSEC}, SPEC~\cite{SPEC} or real \gls{hpc} applications; (ii) simpler but effective performance models can be built on top of workload traces, e.g., \gls{ipc} model and roofline model~\cite{HennessyPatterson12}, enabling the evaluation of the impact of the power management policies; (iii) the power consumption is mainly affected by workload composition, i.e., memory bandwidth and vector/\gls{simd} arithmetic density.

In this paper, we craft a synthetic benchmark (\texttt{Wsynth}) that stresses the control corner cases and consists of maximum power (\texttt{WsynthMax}) and idle power (\texttt{WsynthIdle}) instructions, mixed power instructions (\texttt{WsynthMix}) and lastly instructions with different power densities and fast switching (\texttt{WsynthFast}) to stress the power limiter and the shorter timing constants of the temperature response.

The \textbf{\textit{governor thread}} emulates command dispatching agents such as the operating system running on the \gls{cpu} or the off-chip \gls{bmc} in the motherboard. 
They are tasked to send requirement directives to the controller, such as \gls{dfs} and \gls{dvs} operating points. In the plant simulation presented in this manuscript, the \gls{os} and \gls{bmc} communication is modeled through the \gls{scmi} transport layer. 
Future work will integrate \gls{pldm}/\gls{mctp} transport layers supported by ControlPULP (sec. \ref{subsubsec:io-interf}) in the \gls{hil} framework for \gls{bmc}-related communication.

Finally, the \textbf{\textit{data retriever thread}} collects all the simulation data and periodically dispatches them through the network. It relies on Eclipse Mosquitto as a message broker, which implements the \gls{mqtt} network-based messaging protocol. 
Collected data are fed to \textit{Examon}~\cite{EXAMON}, an open-source~\footnote{\url{https://github.com/EEESlab/examon}} framework that enables performance and energy monitoring of \gls{hpc} systems. \textit{Examon} relies on \textit{Grafana} as interactive data-visualization platform with live-updated dashboards.

\subsection{HIL testing procedure}\label{subsubsec:hil-procedure}

We identify two main phases to carry out the emulation.
In the \textit{system setup} phase, the communication between \gls{ps} and \gls{pl} happens in a conductor-follower fashion. The \gls{ps} drives the \gls{pcs} deployment on the \gls{fpga}, its booting process, and firmware binary flashing. The AMBA AXI4 protocol regulates the data transmission across this communication channel.
To this aim, we generate a complete embedded \gls{smp} Linux system paired with a persistent root file system on top of the four Arm A53 processors with Buildroot~\footnote{\url{https://buildroot.org/}}.

In the subsequent \textit{system emulation} phase, \gls{ps} and \gls{pl} execute the respective routines and communicate \textit{asynchronously} with each other through the shared DRAM memory region.
The lack of an explicit synchronization point between the controller and the simulated plant is inherent to the nature of the control since dynamic thermal and power management involves run-time active control. At the same time, the underlying \gls{mpsoc} varies its workload to meet a specific computational need.

The \textit{system's setup and emulation} phases consist of the following steps:

\begin{enumerate}
    \item Linux \gls{fsbl}: the \gls{pl} is programmed with ControlPULP's \textit{bitfile}, which contains the hardware design information of the controller.
    \item Linux \textit{U-Boot}: Linux kernel boots on the Arm \gls{apu} cores.
    \item ControlPULP is clocked from the \gls{ps}, out of reset, and in idle state. Internal divisions of the external clock are handled within ControlPULP.
    \item The \gls{ps} drives ControlPULP booting process by flashing the L2 SRAM with the control firmware executable (Fig. \ref{fig:cpulp_archi}). 
    \item The \gls{ps} and \gls{pl} start to asynchronously execute the plant simulation and the control firmware routines, respectively, through the shared memory. 
\end{enumerate}

The data are then collected for online dashboard monitoring, as detailed in the previous section.

The key benefit of the proposed methodology is the flexibility gained at the \gls{hw}/\gls{sw} interface. Design space exploration can be carried out on both the hardware controller and the control algorithm, with a short turnaround development time. 
The methodology especially fits integrated (on-chip) control systems validation and co-design due to the native on-chip hardware flexibility offered by the \gls{soc}-\gls{fpga} ecosystem.

\section{Evaluation}\label{sec:results}

\begin{figure*}[t]
    \centering
    \includegraphics[width=0.8\linewidth]{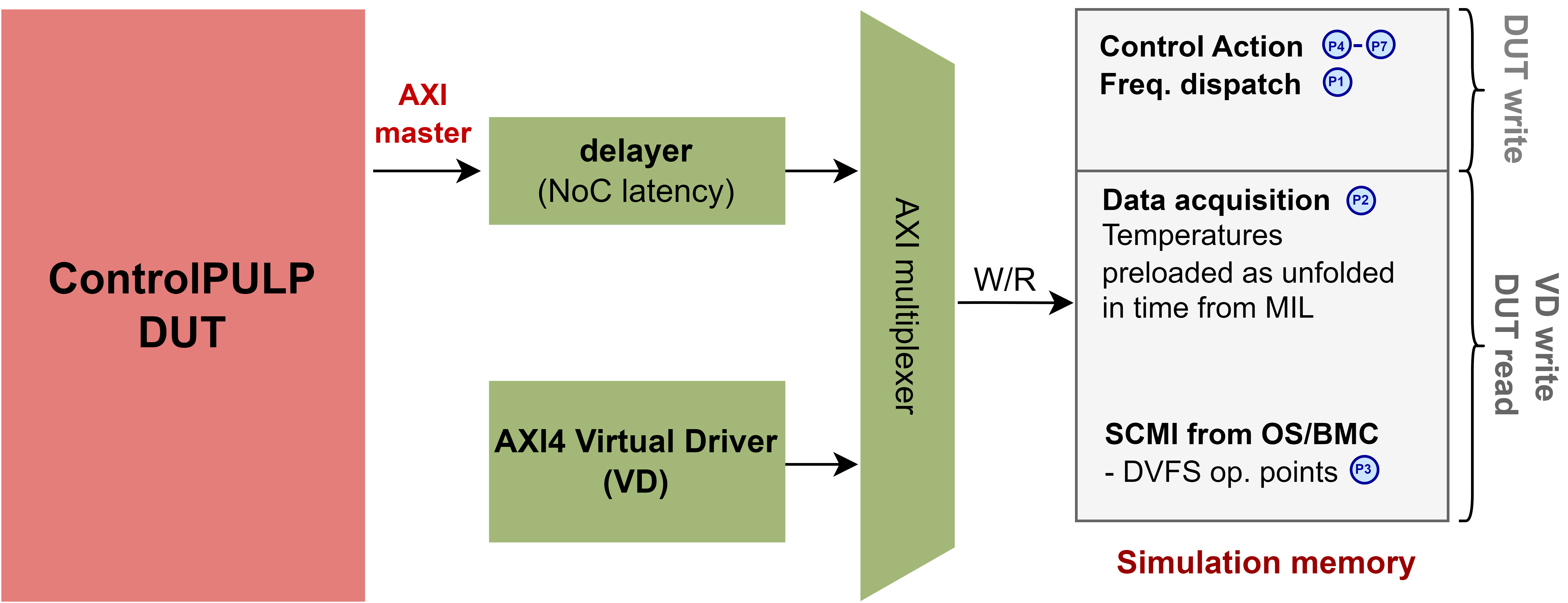}
    \caption{ControlPULP RTL testbench simulation environment}
    \label{fig:fw-tb}
\end{figure*}

In this section, we analyze and characterize both hardware (the ControlPULP platform) and software (the \gls{pcf}) layers:
\begin{itemize}
    \item We break down ControlPULP's post-synthesis area, which represents a small overhead ($<$ 1\%) compared to a modern \gls{hpc} processor die (Sec.~\ref{subsec:cpulp-synth}).

    \item We first evaluate ControlPULP architecture with a cycle-accurate \gls{rtl} testbench environment as depicted in Fig.~\ref{fig:fw-tb}.
    We model the latency of the interconnect network sketched in Fig.~\ref{fig:fw-sketch} by adding a programmable latency to the AXI4 interface.
    In the described test scenario, we first study the parallelization of the \textit{control action} (P4)-(P7) on the cluster (Sec.~\ref{subsec:fw-test}).
    We then characterize \textit{in-band} transfers, namely strided \gls{dma} accesses for data acquisition from \gls{pvt} registers and \gls{clic} interrupt latency with \gls{scmi} command processing (Sec.~\ref{subsec:dma-test}).  
    Finally, we show the overall performance improvement of a single control step when accelerating control tasks in the cluster compared to single-core (Sec.~\ref{subsec:pcf-system-eval}).
    The testbench depicted in Fig. \ref{fig:fw-bars} does not provide the \gls{rtl} description of the surrounding \gls{hpc} processor. Instead, we model the closed-loop with a shared memory region between the \gls{pcs} platform --- the \gls{isut} --- and the system under control. Note that the real-time temperature and telemetry information from the \gls{hpc} processor are pre-computed from a \gls{mil} emulation of the control algorithm executed with a fixed time step and statically stored in the simulation memory as unfolded in time.
    
    \item Since the standalone \gls{rtl} simulation environment fails to provide a near-real-time closed-loop emulation framework for the control algorithm, we rely on the \gls{fpga}-\gls{soc} methodology detailed in Sec. \ref{sec:fpga-map} to analyze the \gls{pcf} \gls{qos} when ControlPULP is mapped on real hardware and compare against a pure \gls{mil} approach to assess the validity of the \gls{hw}/\gls{sw} co-design.

    \item Finally, we show that the proposed \gls{pcf} compares favorably against one of the most well-documented and freely accessible \gls{sota} industrial solutions on the market, IBM's OpenPower (Sec.~\ref{subsec:pcf-control-eval}). The comparison is carried in pure \gls{mil} simulation, being IBM's \gls{pcs} hardware source publicly unavailable.
\end{itemize}

\subsection{Area evaluation}\label{subsec:cpulp-synth}

We synthesize ControlPULP in GlobalFoundries 22FDX FD-SOI technology using Synopsys Design Compiler 2021.06. One gate equivalent (GE) for this technology equals 0.199 $\mu$m$^2$. The design has an overall area of 9.1 MGE, 32\% occupied by the computing cluster accelerator and local L1 memory, and almost half (60\%) by the main L2 memory in the \textit{manager domain}.

We use a system clock frequency of 500 MHz. 
The rationale of the frequency choice is the following: the \gls{pcs} typically belongs to the same clock domain of the \gls{cpu}'s \textit{uncore} region. While the \textit{core} region hosting the application-class \glspl{pe} runs at frequencies in the order of 1-3GHz, the \textit{uncore} is clocked at lower frequencies, in the order of hundreds of MHz, with modern architectures and technology nodes.

The target controlled-system die area is assumed comparable to other commercials, multi-core ($>64$) server-class processors such as~\cite{GRAVITON}, about 460 mm$^2$. By correlating the gate-equivalent count of the \gls{hpc} \gls{cpu} die in the same technology node of this work, ControlPULP would still represent about 0.5\% of the available die area\footnote{This has to be considered a first approximation, since it compares post-synthesis results with publicly available data of a modern \gls{hpc} die, nowadays manufactured in a more advanced technology node.}. 
This first-order estimation makes the design choice of a parallel \gls{pcs} valuable since its capabilities are much increased, while the silicon area cost remains negligible within a high-performance processor die.

\begin{table}[t]
    \caption{ControlPULP post-synthesis area breakdown on GF22FDX technology.}
    \begin{center}
    \renewcommand{\arraystretch}{1.2} 
    \begin{tabular}{cccc}
        \hline
        \textbf{Unit} & \textbf{Area [mm$^2$]} & \textbf{Area [kGE]} & \textbf{Percentage [\%]} \\
        \hline
        Cluster domain  & 0.467 & 2336.7 & 25.5\\
        Manager domain   & 0.135 & 675.9 & 7.4\\
        L1 SRAM & 0.119 & 595.7 & 6.5\\
        L2 SRAM  & 1.108 & 5542.1 & 60.6\\
        \textbf{Total} & \textbf{1.830} & \textbf{9150.3} & \textbf{100} \\ 
        \hline
        \end{tabular}
    \label{tab:ps-area}
    \end{center}
\end{table}

\subsection{Firmware \textit{control action}}\label{subsec:fw-test}

In the following, we analyze the execution of the \gls{pcf} phases (P4)-(P7) on the multi-core cluster accelerator. We enforce power capping (\textit{alpha} reduction~\cite{gap8-hil}) to evaluate each computational phase fairly. 
Each cluster core is responsible for a subset of the controlled \glspl{pe}.
The parallelization is implemented as a fork-join process where the workload is statically distributed among the workers. In ControlPULP, the construct is implemented through a per-worker \texttt{thread\_id} $\in [0:N_{workers}-1] $ and an equally distributed \texttt{chunk\_size} where
$   \texttt{chunk\_size} = \displaystyle\frac{N_{ctrl\_cores}}{N_{workers}}$. We are interested in extracting performance figures for the \textit{control action} in a single periodic step $n$. We execute the \gls{pcf} for $S$ steps to amortize the effect of the initially cold instruction cache.
Finally, we perform the arithmetic mean over $S$ to get the mean absolute execution time for each (P4)-(P7) phase.

We report the execution time $\tau_0$ and the multi-core speedup ($\frac{\tau_{0,single}}{\tau_{0,multi}}$) at varying number of controlled cores N$_{cc}$ for each \gls{pcf} phase in Figs.~\ref{fig:fw-bars} and~\ref{fig:fw-speedup} respectively. The total speedup of the full \textit{control action} at fixed N$_{cc}$ is the geometric mean over the speedups of each phase.  
In our use case of 72 controlled \glspl{pe}, ControlPULP executes the \textit{control action} 5.5x faster than in single-core configuration, reaching 6.5x with 296 controlled cores. 

We make the following observations.
First, multi-core speedup scales with the number of controlled cores due to the increased workload and is affected by the workload characteristics of each phase. 
Second, the \textit{control action} is not a fully computational step. In fact, instruction branching associated with power and frequency bounds checks per core introduces additional load/store stalls due to data access contention in a multi-core configuration.
Finally, the computational body of (P6) and (P7) can be separated into independent parallel tasks and is thus an embarrassingly parallel problem. 
Instead, (P4) and (P5) show dependency across the values computed by the workers in the form of reduction sums, i.e., in (P4) to calculate the total power of the \gls{cpu} and (P5) to calculate a normalization base for \textit{alpha} power capping~\cite{gap8-hil} and again the total \gls{cpu} power. 
When a reduction sum is needed, we use a hardware barrier to synchronize the threads and join the concurrent execution on the cluster master core (core 0), which carries out the reduction. 

As discussed in the analysis above, the increased parallel compute capability of handling the control's computational workload, paired with the general purpose nature of the accelerator, enable us to (i) improve the control performances with more advanced algorithms and (ii) be fully flexible when designing the control algorithm.

\subsection{\textit{in-band} Services}\label{subsec:dma-test}

\subsubsection{PVT sensors}
\begin{figure*}[t]
	\centering
	\subfloat[\label{fig:fw-bars}]{\includegraphics[width=0.45\columnwidth]{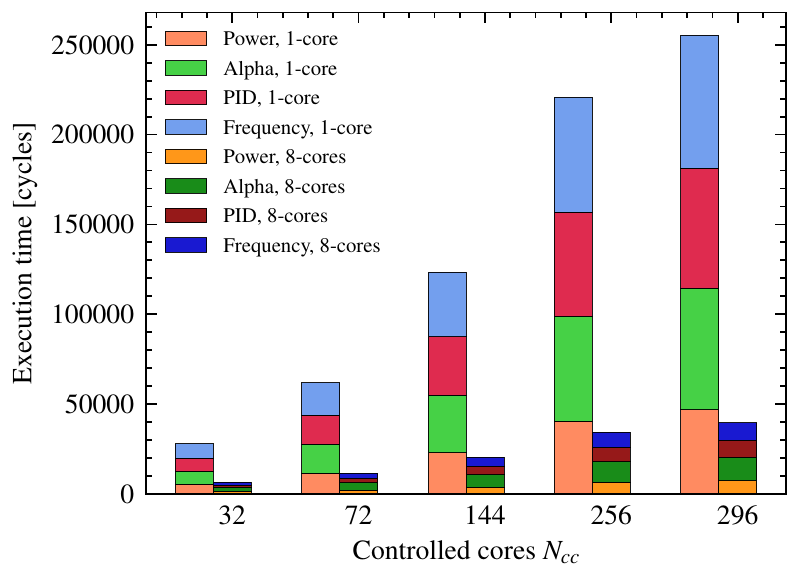}}
	\qquad
	\subfloat[\label{fig:fw-speedup}]{\includegraphics[width=0.415\columnwidth]{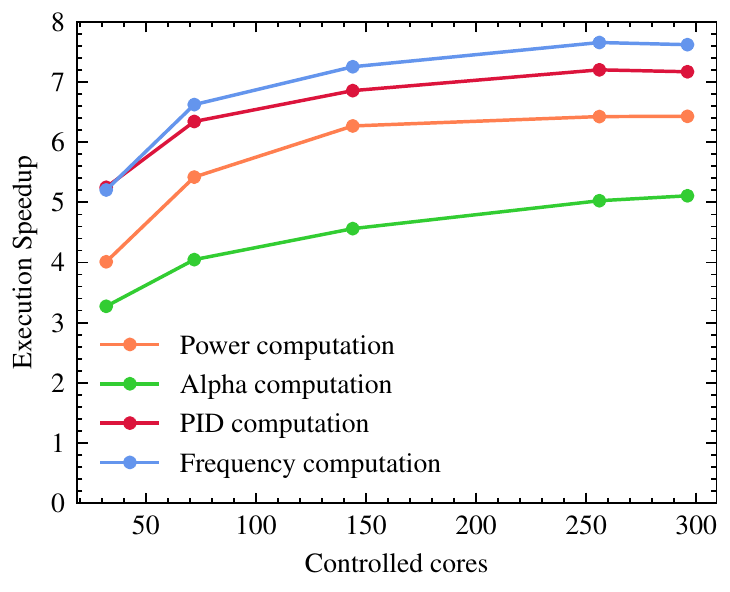}}
	\qquad
    \subfloat[\label{fig:dma}]{\includegraphics[width=0.42\columnwidth]{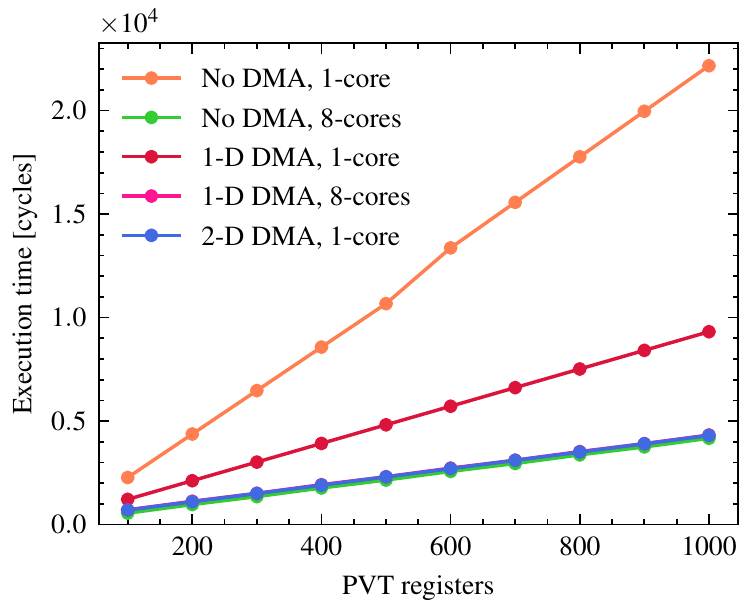}}
    \qquad
    \subfloat[\label{fig:scmi_irq_basic_cycles}]{\includegraphics[width=0.44\columnwidth]{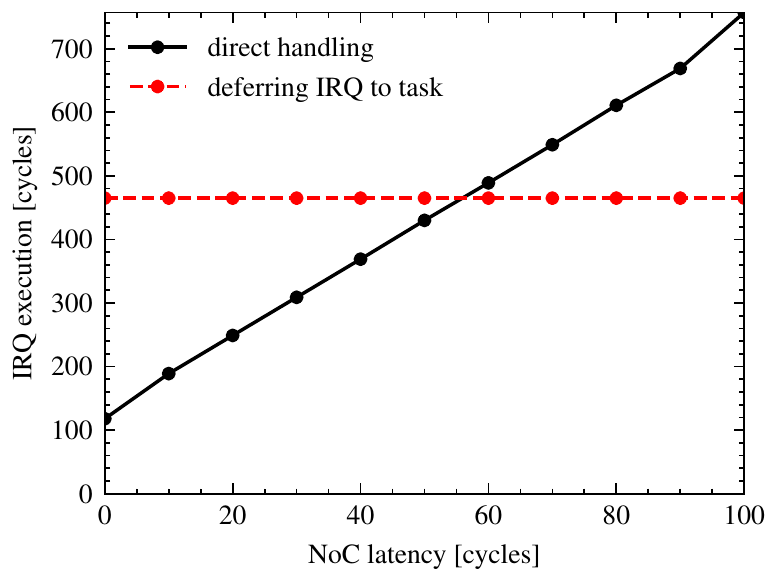}}
    
    \caption{\textbf{(a)-(b)} Firmware \textit{control action}, execution time, and speedup comparison between single-core (\textit{manager domain}) and 8-cores (\textit{cluster domain}).
    \textbf{(c)} \textit{in-band} data acquisition from simulated \gls{pvt} registers, execution time without and with \gls{dma} in 1-D and 2-D configurations.
    \textbf{(d)} Execution time in the interrupt handler from the interrupt edge to its completion with a basic \gls{scmi} message at varying interconnect network access latency to the mailbox}

\end{figure*}

To assess \textit{in-band} services involving \gls{pvt} physical sensors --- phases (P1) and (P2) ---, we measure the transfer time required for reading data bursts on the AXI4 master bus with the \gls{soc} timer. 
The exploration is three-fold: (i) direct data gathering from the ControlPULP cluster's cores, (ii) data gathering by offloading the transfers to the \gls{dma} in 1-D configuration, and (iii) \gls{dma} offload in 2-D configuration~\cite{DMA}. For (i) and (ii), we investigate the data collection on either 1-core or 8-cores configurations.  The address range is equally distributed among the issuing cores in the latter scenario. In (iii), one core performs the read operation to highlight the advantages of offloading a single, large transfer with non-contiguous but uniformly spaced addresses to the \gls{dma}, which increases the addresses by the selected stride. 
This configuration becomes important when atomically gathering \gls{pvt} information from equally spaced address locations (\gls{hpc} \glspl{pe}) with only one transfer request. 
As in Sec. \ref{subsec:fw-test}, we use synchronization barriers to coordinate the eight cores.
Fig. \ref{fig:dma} reports the execution time $\tau_1$ required for data movement when reading from up to 1000 \gls{pvt} registers (4B each), an estimate bound given the number of \glspl{pe} and the information needed from them (P, V, T, i.e., $\ge\>3$, lower bound).
Fig. \ref{fig:dma} shows that the best \gls{dma}-based transfers assuming 1000 \gls{pvt} registers (2-D) are 5.3x faster than single-core direct data gathering.

\subsubsection{SCMI and interrupt latency}\label{subsec:scmi-test}

\begin{table}[t]
    \caption{Interrupt latency from interrupt edge to the first instruction in the interrupt handler as number of cycles.}
    \begin{center}
    \renewcommand{\arraystretch}{1.2} 
    \begin{tabular}{ccc}
        \hline
        \textbf{Location} & \textbf{Increment [cycles]} & \textbf{Sum [cycles]} \\
        \hline
        CLIC input to output  & 1 & 1 \\
        CLIC output to core (handshake)  & 2 & 3 \\
        Claim interrupt  & 1 & 4 \\
        Jump in vector table to CLIC handler   & 2 & 6 \\
        Save caller save regs (\texttt{addi} + 15 regs) & 17 & 23 \\
        Compute and load CLIC handler address & 5 & 28 \\
        Jump to CLIC handler address          & 2 & 30 \\
        \textbf{Summary}                      & - & \textbf{30} \\
        \hline
        \end{tabular}
    \label{tab:plic_int_latency}
    \end{center}
 \end{table}

An interrupt-driven (doorbell) transport regulates the communication of \gls{dvfs} operating points in the agent-platform direction.
Table \ref{tab:plic_int_latency} gives an overview of the overall CLIC interrupt latency measured as the number of cycles from the triggering edge in the CLIC to the ISR Handler's first instruction. The configuration of interrupt level, priority, and threshold configuration is handled with memory-mapped and CSR accesses to the \gls{clic} register file and the manager core, respectively. This leads to a lower programming latency than software-driven approaches required in standard RISC-V \gls{plic} or \gls{clint} interrupt controllers.


With a working frequency of 500MHz, the interrupt latency of one \gls{scmi} command coming from the \gls{os} \gls{hlc} controller governor is negligible compared to the period of the \gls{pfct} that executes every 500$\mu$s, namely 0.01\%.

Analogously, the context switch time needed to preempt the \gls{pfct} with the \gls{pvct}, which runs 4 times faster (every 125$\mu$s), during the execution of a \gls{pcf} step is 0.08\% the available time period of the \gls{pfct}, more than enough for the two tasks to coexists while executing their respective policies.

It is essential to notice that the latency of the interconnect \gls{noc} between ControlPULP and the mailboxes located in the die (Figs. \ref{fig:fw-sketch}) has a significant impact on the load/store access times, thus the time spent in the ISR, which grows with the interconnect delay size. 
We show this effect by emulating the shared mailboxes as well as the \gls{noc} latency in the \gls{rtl} testbench environment (Fig.~\ref{fig:fw-tb}). The black line in Fig. \ref{fig:scmi_irq_basic_cycles} reports the execution time for directly decoding and responding to a sample \gls{scmi} command (\textit{Base Protocol}, \texttt{protocol\_id = 0x10}, \texttt{message\_id = 0x0}~\cite{SCMI}) in the ISR when an external simulated driver rings a doorbell to the \gls{pcs}. The figure reveals that the time spent in the ISR linearly increases with the \gls{noc} latency.
We tackle the impact of \gls{noc} latency by deferring pending interrupts as they are triggered, thus keeping the ISR time short and insensitive to the \gls{cpu} interconnect network delay, with the FreeRTOS timer API \texttt{xTimerPendFunctionCallFromISR()}.
From the red line in Fig. \ref{fig:scmi_irq_basic_cycles}, we see that deferring interrupt handling to a task is preferable over direct handling, as it is network-latency insensitive for realistic \gls{noc} latencies larger than 50 cycles.
Other existing solutions, such as Arm SCP firmware, propose a bespoke \textit{Deferred Response Architecture}~\footnotemark[\getrefnumber{scp_open}] to mark selected requests coming from an agent as pending and defer the platform response. We instead rely on a trusted scheduler that decouples \gls{os} and \gls{pcf} driver APIs, improving flexibility and portability. 


\subsection{System-level PCF step evaluation}\label{subsec:pcf-system-eval}

We finalize the standalone evaluations of ControlPULP's architectural features from the previous sections with the overall \gls{pfct} step cycle count comparison between accelerator-enhanced and single-core configurations, reported in Table \ref{tab:overall_gain} in the case of 72 controlled cores.
Table \ref{tab:overall_gain} shows a breakdown of the required actions.

The total execution time differs in the two execution models. In the single-core case, we execute sequentially with less overhead from data movement ($ T_{single} = \tau_0 + \tau_1 $).
In the multi-core case, ($ T_{multi} = max(\tau_0, \tau_1) + \sum_{i=2}^{5} \tau_i $) we (i) execute the computation $\tau_0$ and data acquisition $\tau_1$ at step $n$ concurrently, (ii) rely on $\tau_{0,multi} \ll \tau_{0,single}$, and (iii) introduce an overhead due to additional data movement involving L1 and L2 for data telemetry between \textit{manager} and \textit{cluster domains} during the \gls{pfct}.  

\begin{figure*}[t]
    \centering
    \includegraphics[width=0.7\linewidth]{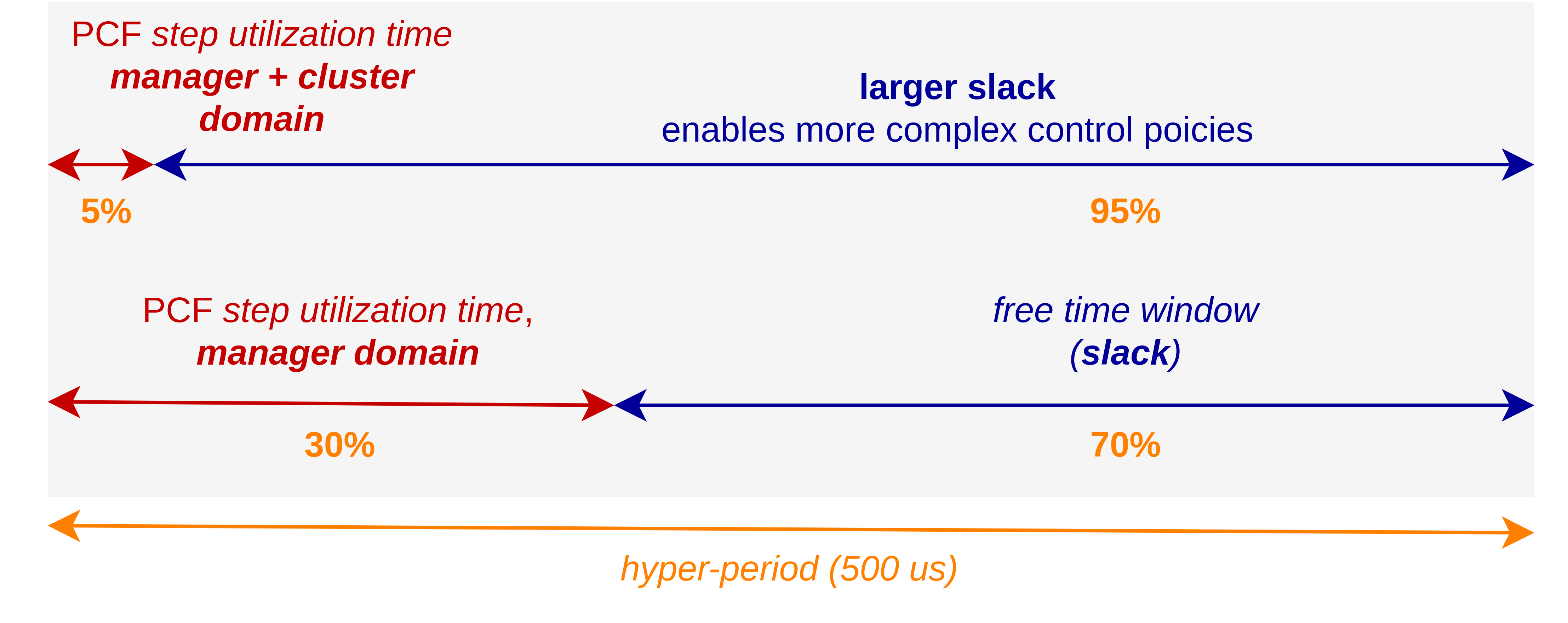}
    \caption{Benefits of a performant parallel \gls{pcs} on the control problem, namely, room for more computational-intensive control policies from the concurrent acceleration. The figure shows that the hyper-period free time window (\textit{slack}) increases to almost 95\% when the policy is accelerated with the \textit{cluster domain}}
    \label{fig:hyperperiod}
\end{figure*}

Overall, multi-core execution achieves a 4.9x speedup over the single-core configuration. 
Provided a fixed \textit{hyper-period}, i.e., the least common multiple of the control tasks’ periods~\cite{HYPERPERIOD}, which in this work equals the \gls{pfct} step period of 500$\mu s$, Fig. \ref{fig:hyperperiod} shows the benefits of a programmable accelerator on the control policy time scale.
Assuming a working frequency of 500MHz, and considering the interrupt latency and task preemption context switch time from the scheduler negligible as from  Sec. \ref{subsec:scmi-test}, single-core execution time already leaves a free time window (\textit{slack}) of about 70\% the hyper-period with the reactive control algorithm implemented in this work. 
Cluster-based acceleration significantly raises the free time window to about 95\% the \gls{pfct} period.
This means the acceleration reduces the utilization time to complete the control task within the deadline. 
Furthermore, the more embarrassingly parallel the control problem, the more the concurrent speedup, hence the benefits of the acceleration.

\revision{Other effects in the plant can impact the control loop period in the control scenario presented in this work: (i) \gls{pll} lock time and voltage regulator response time, (ii) communication latency to sensors/actuators, and (iii) model dynamics evolution.
\gls{pll} lock time and voltage regulator response time are in the order of microseconds. Similarly, communication latency depends on the interconnect network of the \gls{cpu}. The \gls{pcf} computes the frequencies/voltages at time step \textit{n} and applies them at time step \textit{n+1}. Therefore, latency issues - if any - due to the effects of (i) and (ii) can be addressed by starting early the application of the set points during time step \textit{n}, provided the control period has enough slack time after the computation phase.
Model dynamics (iii) are tightly coupled with the system in use and can be broken down into workload, power, and temperature effects.
Temperature variations are in the order of magnitude of \textit{ms}.  
Workload and power can potentially change at each clock cycle. Arm proposes the Maximum Power Mitigation Mechanism (MPMM) for mitigating and limiting the execution of high-activity events, such as vector instructions. The management of this mechanism is left to the power controller. This could be a potential extension to ControlPULP as well.

The $500 \mu s$ hyperperiod used in this work is shorter than the fastest dynamics in terms of temperature, but longer than the workload and power variations. 
The latter case represents a worst-case scenario to guarantee safe margins on the power constraint in actual \gls{hpc} benchmarks.}

\begin{table}[t]
    \caption{Execution time $T$ of a \gls{pfct} step, single-core and cluster configurations. \gls{scmi} commands exchange and off-die transfers, handled by the \gls{soc} \textit{manager core}, are not included in the comparison since they are a shared overhead.}
    \begin{center}
    \renewcommand{\arraystretch}{1.2} 
    \resizebox{\columnwidth}{!}{
    \begin{tabular}{ccccc}
        \hline
        \textbf{Firmware phase} & \textbf{Time step} & \multicolumn{2}{c}{\textbf{Execution time [cycles]}} & \textbf{Speedup} \\
        & & \textbf{1-core}& \textbf{Multi-core} &\\
        \hline
        \textit{control action} (P4)-(P7) & $\tau_0$ & 61867 & 11372 & 5.5x\\
        \textit{in-band} transfers (P1),(P2),(P3) & $\tau_1$ & 5463 & 3523 (DMA) & 1.6x\\
        Offload to the Cluster & $\tau_2$  & - & 389 & -\\
        L2 - L1 transfers & $\tau_3$ & - & 434 (DMA) & -\\
        L1 - L2 transfers  & $\tau_4$  & - & 872 (DMA) & - \\
        Return from Cluster & $\tau_5$ & - & 574 & - \\
        \textbf{Step total time} & T & 67330 & 13641 & \textbf{4.9x}\\
        \hline
        \end{tabular}
        }
    \label{tab:overall_gain}
    \end{center}
\end{table}

\subsection{Control-level PCF evaluation}\label{subsec:pcf-control-eval}

We refer to the \gls{pcf} \gls{qos} as an indicator of the control policy functional correctness (\gls{hw} and \gls{sw}) when the simulated \gls{mpsoc} is assigned a certain workload. 
In this section, we present the results of this analysis by leveraging the \gls{fpga}-\gls{soc} closed-loop framework, the thermal and power model, and the synthetic workload introduced in sec. \ref{subsubsec:hil-procedure}. We subsequently complete the \gls{qos} exploration with a comparison against the IBM OpenPower open-source control algorithm.

\subsubsection{Standalone QoS evaluation on the HIL FPGA-SoC framework}

Provided a workload assigned to the simulated \gls{mpsoc}, the \gls{hil} framework enables assessing the evolution in time of critical features of the control algorithm at the granularity of each controlled core (Sec. \ref{subsec:fw-archi}): power/thermal capping, workload-aware control, and frequency/voltage set-point tracking as a response to \gls{os} and \gls{bmc} \gls{dvfs} commands.

\begin{figure*}[t]
	\centering
	\subfloat[Thermal capping on a subset of \glspl{pe}.]{\includegraphics[width=0.5\linewidth]{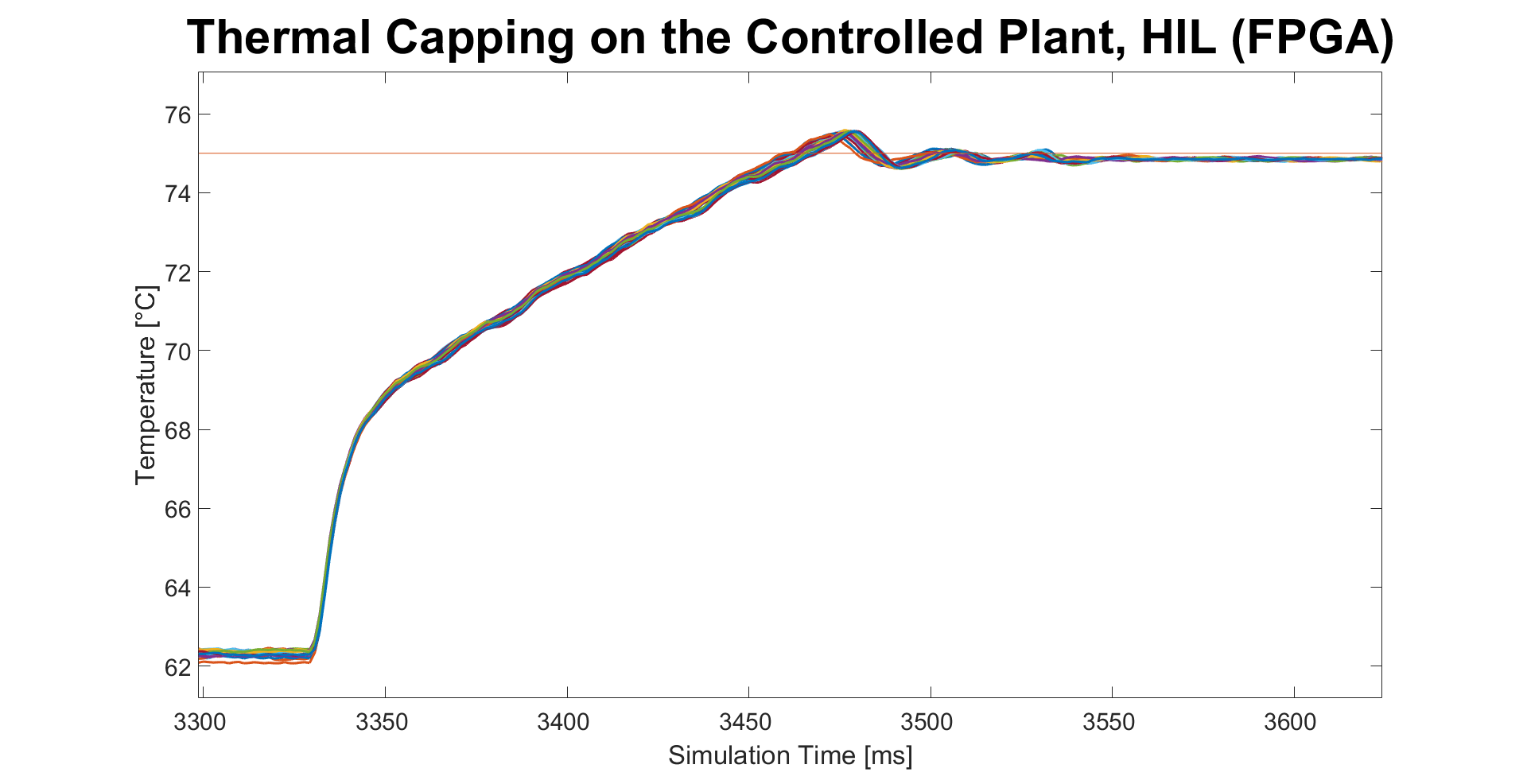}
	\label{fig:hil-thcap}}
	\subfloat[Total power consumption and capping action.]{\includegraphics[width=0.5\linewidth]{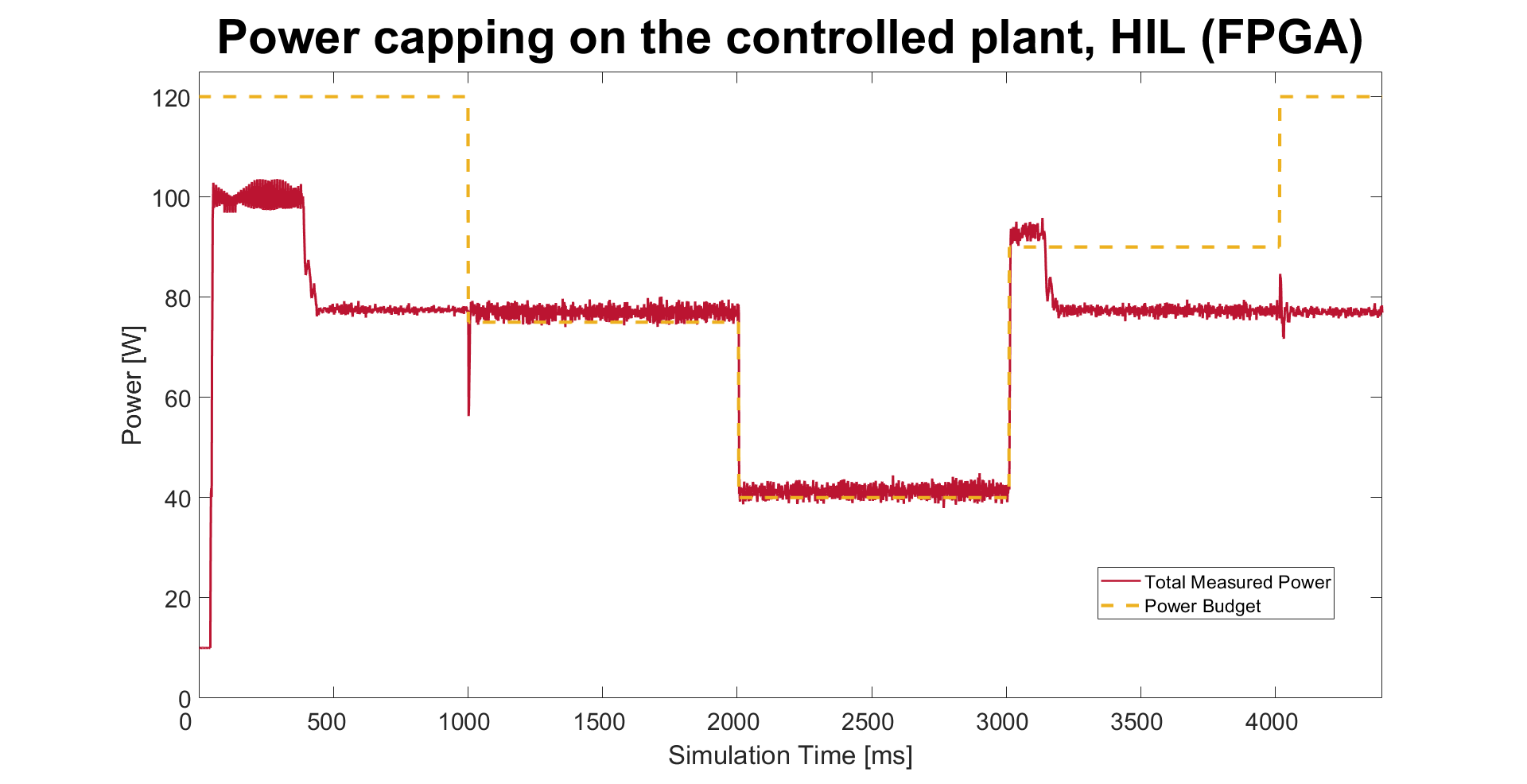}
	\label{fig:hil-pwcap}}
	
	\subfloat[Dynamic frequency scaling on a subset of \glspl{pe}.]{\includegraphics[width=0.5\linewidth]{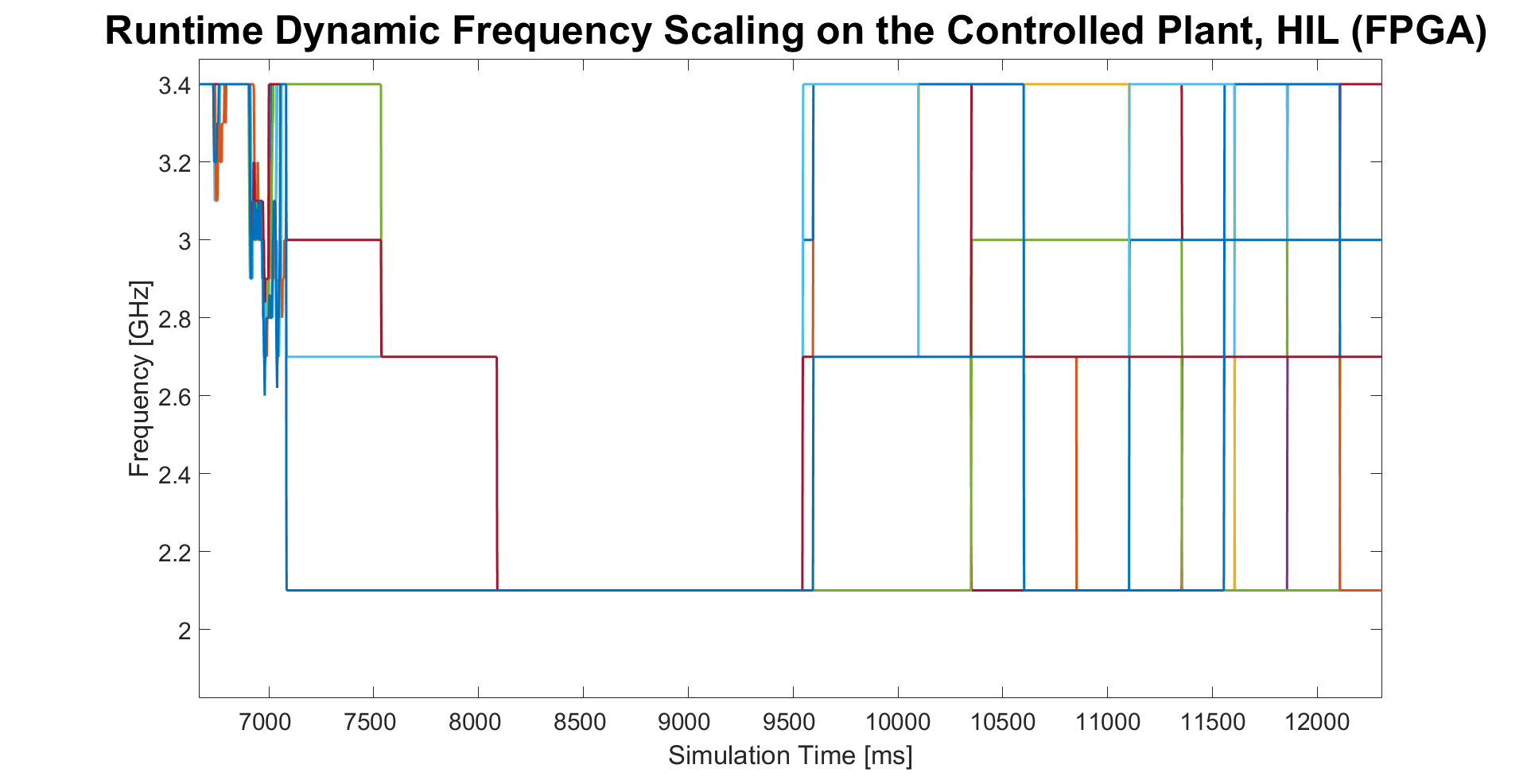}
	\label{fig:hil-dfs}}
	\subfloat[Dynamic voltage scaling on a subset of \glspl{pe}.]{\includegraphics[width=0.5\linewidth]{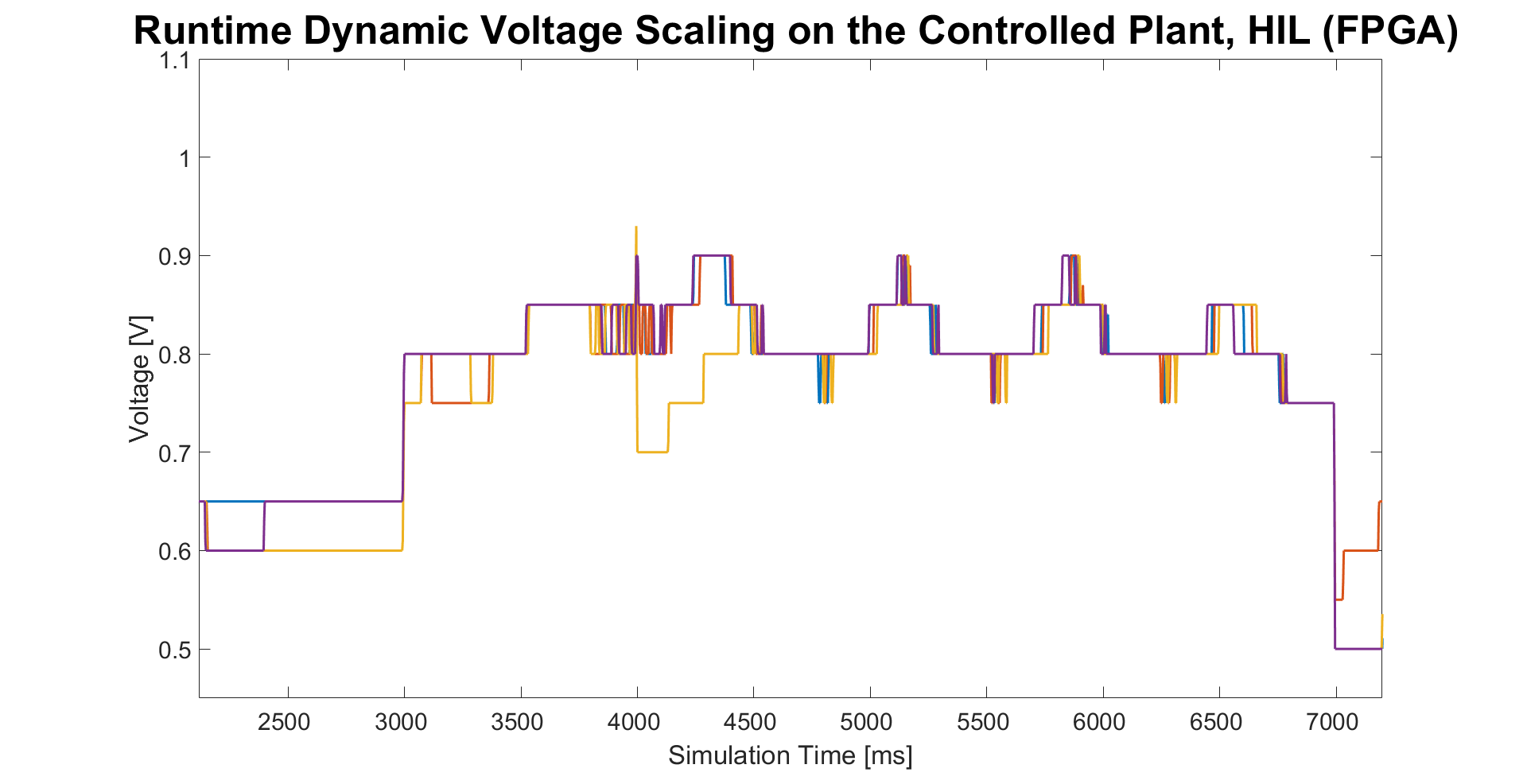}
	\label{fig:hil-dvs}}
		
    \caption{Thermal, power capping and \gls{dvfs} emulation on the \gls{fpga}-\gls{soc} \gls{hil} framework with \texttt{Wsynth}. 
    Requests coming from the \glspl{hlc} (\gls{os} or \gls{bmc}) governors, such as the target frequency and power budget, are processed by the reactive control, phases (P6)-(P7) described in Sec. \ref{subsec:fw-archi}.
    The number of controlled cores is 72}
\end{figure*}

Figs. \ref{fig:hil-dfs} and \ref{fig:hil-dvs} show the frequency and voltage scaling enforced when the controller's \gls{scmi} mailboxes are notified \gls{dvfs} operating point commands from the plant's \textit{governor thread} running on the \gls{ps} on a per-\gls{pe} basis (Sec. \ref{subsubsec:hil-plant_sim}). 
The \gls{scmi} agent dispatches the commands according to the executed workload; interrupts processing is deferred by the FreeRTOS scheduler and committed by ControlPULP's interrupt controller. Once registered, the governor's directive is processed by the \gls{pfct} in phases (P3)-(P7) and returns the reduced frequency as computed in the \textit{cluster domain}.
Fig. \ref{fig:hil-thcap} and \ref{fig:hil-pwcap} show per-core thermal capping and the total power consumption and capping in action on the controlled plant during the execution of \texttt{Wsynth}, respectively. 
The maximum thermal limit from Fig. \ref{fig:hil-thcap}, specific for each \gls{pe}, is assumed to be 85 \textcelsius. The orange line represents instead an additional thermal threshold required to stabilize the temperature with a safety margin in case of overshoots during the \gls{pid}.
Analogously, in Fig. \ref{fig:hil-pwcap}, the total power consumption of the system is bound by the power budget imposed through \gls{scmi} from the \glspl{hlc}.

\begin{figure*}[t]
	\centering
    \subfloat[Evolution in time  of the average frequency in the simulated tile.]{\includegraphics[width=\linewidth]{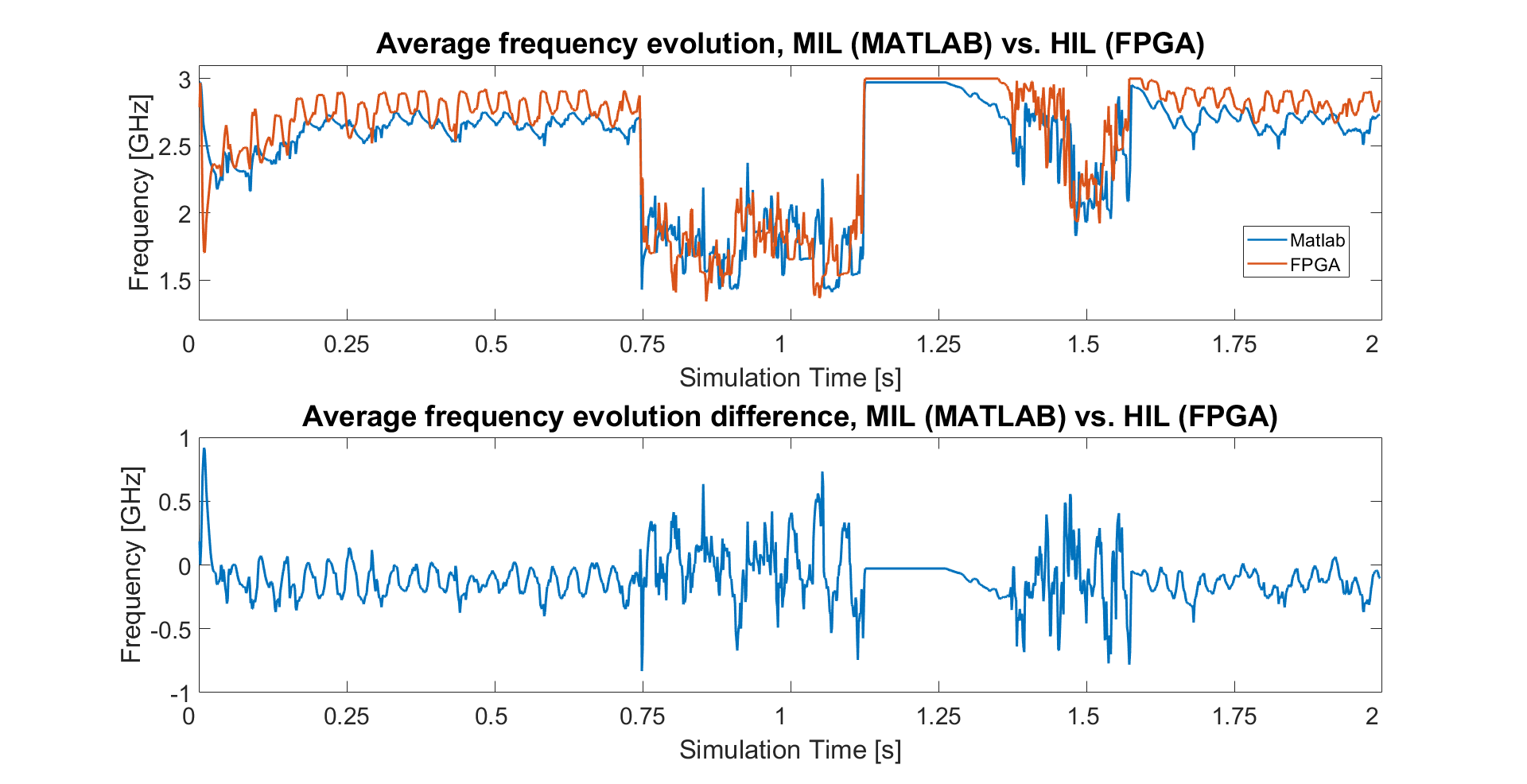}
    \label{fig:hil-mil-w5}}
    
    \subfloat[Evolution in time of the average consumed power in the simulated tile. The dashed yellow line represents power budget directives dictated by the \glspl{hlc}.]{\includegraphics[width=\linewidth]{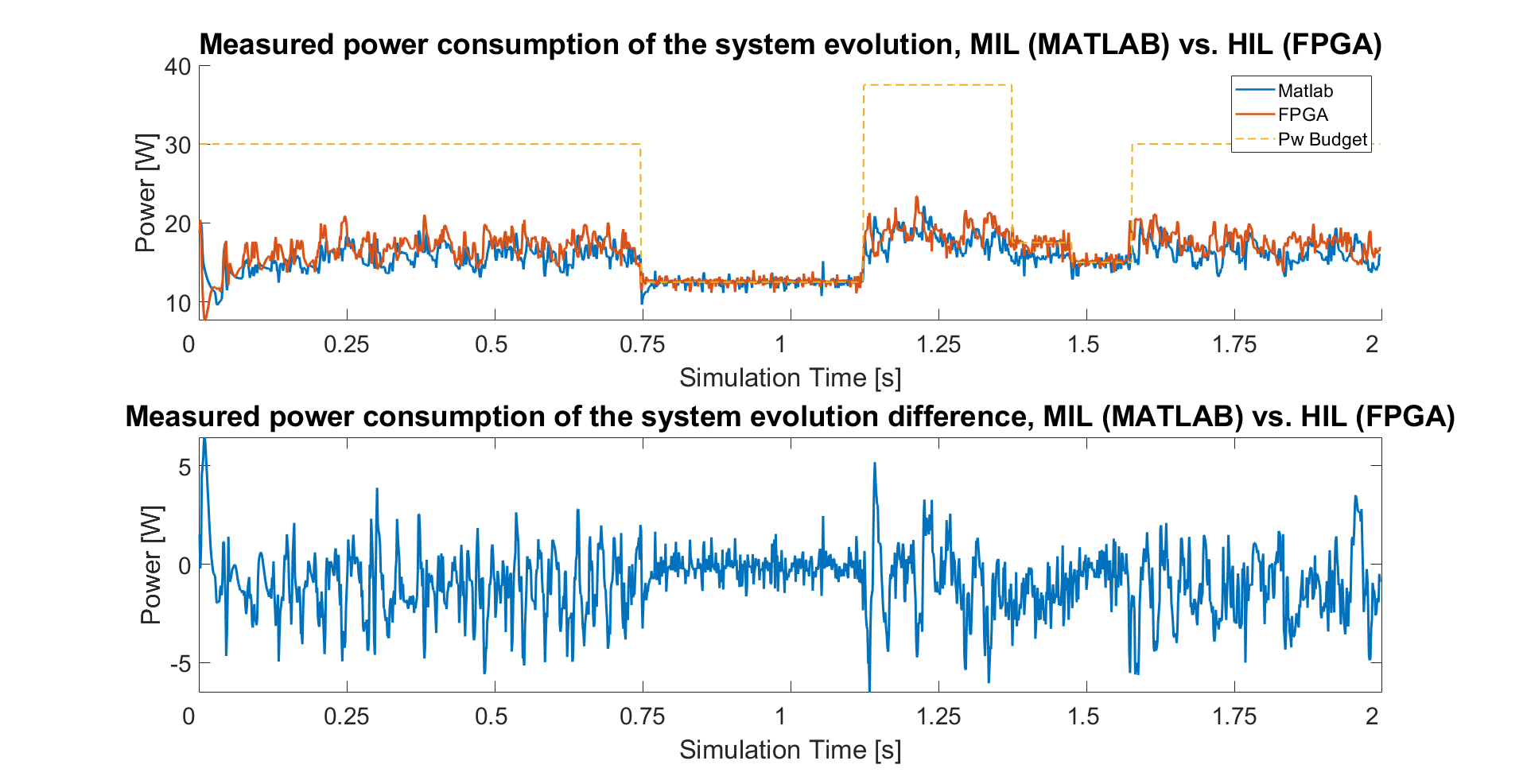}
    \label{fig:hil-mil-w6}}
    
    \caption{\gls{hil} and \gls{mil} comparison when \texttt{Wsynth} is assigned to the controlled system. The emulation assumes a floorplan with a tile of 9 cores and runs for $2s$ to align with MATLAB runtime}
\end{figure*}

The \gls{hil} simulation is further compared with the software-equivalent model-based closed-loop from MATLAB Simulink, the first phase of the control algorithm design.
Due to MATLAB Simulink runtime execution, we restrict the floorplan of the controlled 72-core \gls{cpu} to a tile of 9 cores, let the \gls{pvct} apply a constant voltage $V_{fixed} = 0.75V$ and fix the simulation duration to $2s$.
Fig. \ref{fig:hil-mil-w5} shows the average frequency evolution in time for the controlled cores in the tile. The discrepancy between the outcome of the two simulations has multiple reasons: (i) different data collection methodology: MATLAB records data at the exact time, while the \gls{fpga} simulation captures the information at a non-deterministic sequence of instants in the simulation interval due to its real-time characteristic, (ii) different resolution, and (iii) uncertainties and non-deterministic control delays introduced in the \gls{fpga} emulation and challenging to replicate in the \gls{mil} framework.
Fig. \ref{fig:hil-mil-w6} shows the measured consumed power of the assigned workload under power budget constraints from the \glspl{hlc}, represented with a dashed yellow line.
Overall, the \gls{hil}-based emulation gives comparable results when validated against the software-equivalent \gls{mil}. Albeit power spikes due to the discrepancies described above, \gls{dvfs} tracking achieves a mean deviation within $3\%$ the system's \gls{tdp} (120W in the emulation), more than acceptable for the assessment.

\subsubsection{Comparative QoS evaluation with SOTA}
We cross-benchmark the \gls{pcf} with the IBM OpenPOWER (Sec. \ref{sec:related_work}). 
We model IBM's control action, i.e., the \textit{Voting Box Control} described in Sec. \ref{sec:related_work}, excluding a few architecture-specific features, and the two-layer \gls{pcf} control described in this work with MATLAB Simulink to enable a fair and hardware-agnostic comparison. 
The \gls{pid}-like coefficients of the IBM control are adapted to the \gls{hpc} chip model power and thermal characteristics.
\texttt{Wsynth} (Sec. \ref{subsubsec:hil-plant_sim}) is distributed as follows: \textit{core 1}-\textit{core 3} and \textit{core 2}-\textit{core 4} pairs are assigned \texttt{WsynthMax} and \texttt{WsynthIdle} respectively. \textit{Core 5}, \textit{core 6}, and \textit{core 9} execute \texttt{WsynthMix} while \textit{core 7} and \textit{core 8} are exposed to \texttt{WsynthFast}.
We rely again on constant voltage $V_{fixed} = 0.75V$ and do not consider overhead nor delays in the PLLs and \glspl{vrm} operating point transitions. The power budget is changed five times during the simulation to stress all the elements of the control action. The simulation runs for $2s$ on a tile of 9 cores.

\begin{figure*}[t]
	\centering
	\includegraphics[width=0.7\columnwidth]{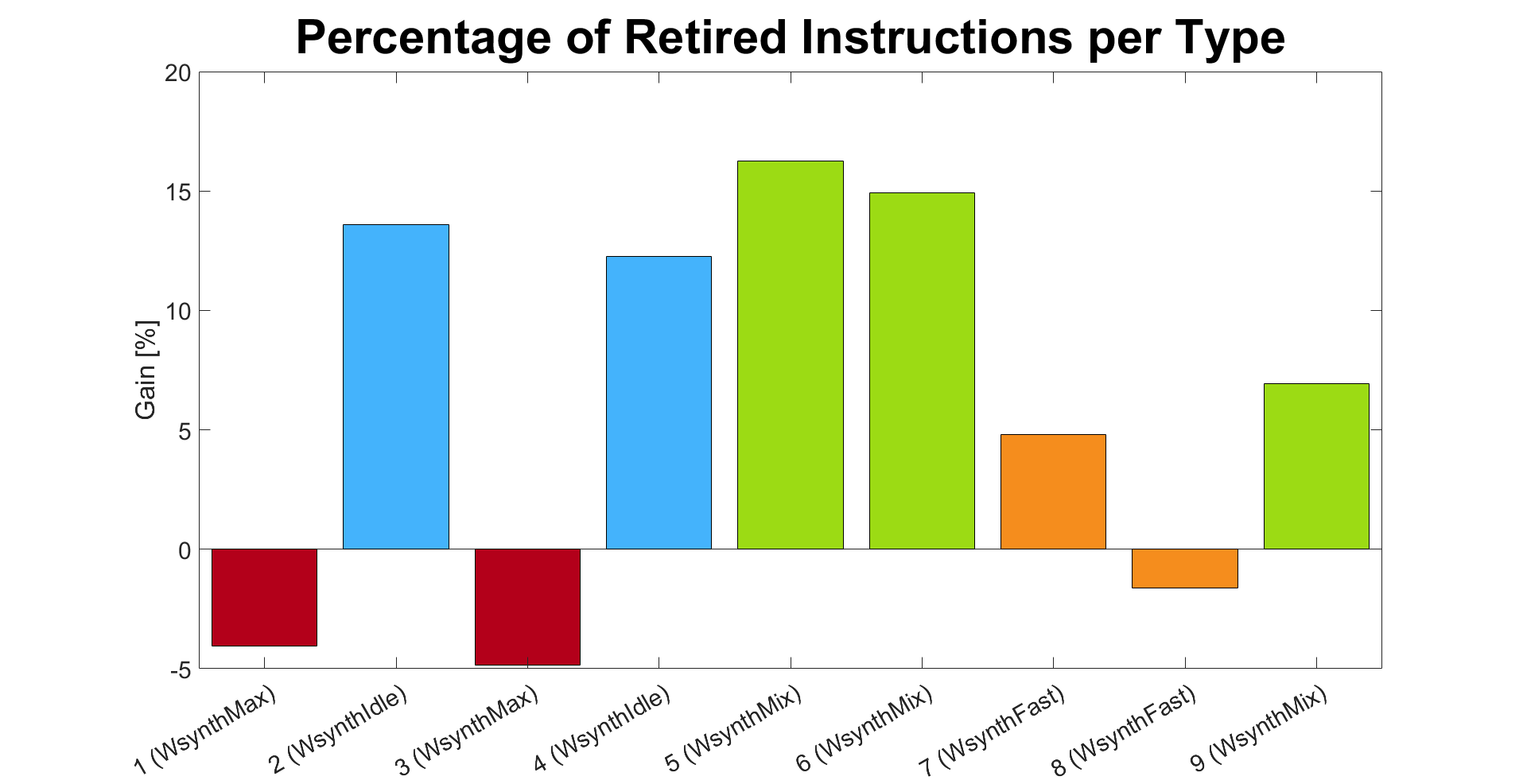}
	\caption{\label{fig:comp2_perf} Comparison between the modified IBM OpenPOWER control with per-core temperature \gls{pid} for frequency reduction and the original IBM OpenPOWER control. The simulation time is $2s$}
\end{figure*}

First, we show that a controller with a multi-core cluster able to deliver higher computational power is beneficial to the performances of the \gls{hpc} chip. We compare the IBM control and a version of it with a per-core temperature \gls{pid} for frequency reduction. In fact, as from Sec. \ref{sec:related_work}, the IBM control policy considers the maximum temperature among the \glspl{pe} when applying frequency reduction. We conjecture that this limitation is enforced by the limited control policy complexity that can be handled by IBM's \gls{occ}. Conversely, ControlPULP enables fine-grained frequency reduction on a per-core temperature granularity.
The performance is shown in Fig.~\ref{fig:comp2_perf}. The number of retired instructions indicates the execution time achieved by the workload at the end of the simulation: the more retired instructions, the faster the workload, meaning a more efficient control.

While using only one temperature for the whole tile results in an average performance reduction per core of $5\%$, cores executing high-power instructions (\textit{core 1} and \textit{core 3}) receive a performance increase of $4\%$ and $5\%$ respectively. In fact, being the frequency reduction based on the hotter cores and thus a shared penalty, neighboring cores get colder, and other cores consume less power during power capping phases, leaving more power available to boost performances of \textit{core 1} and \textit{core 3}.

\begin{figure*}[ht]
	\centering
	\includegraphics[width=0.7\columnwidth]{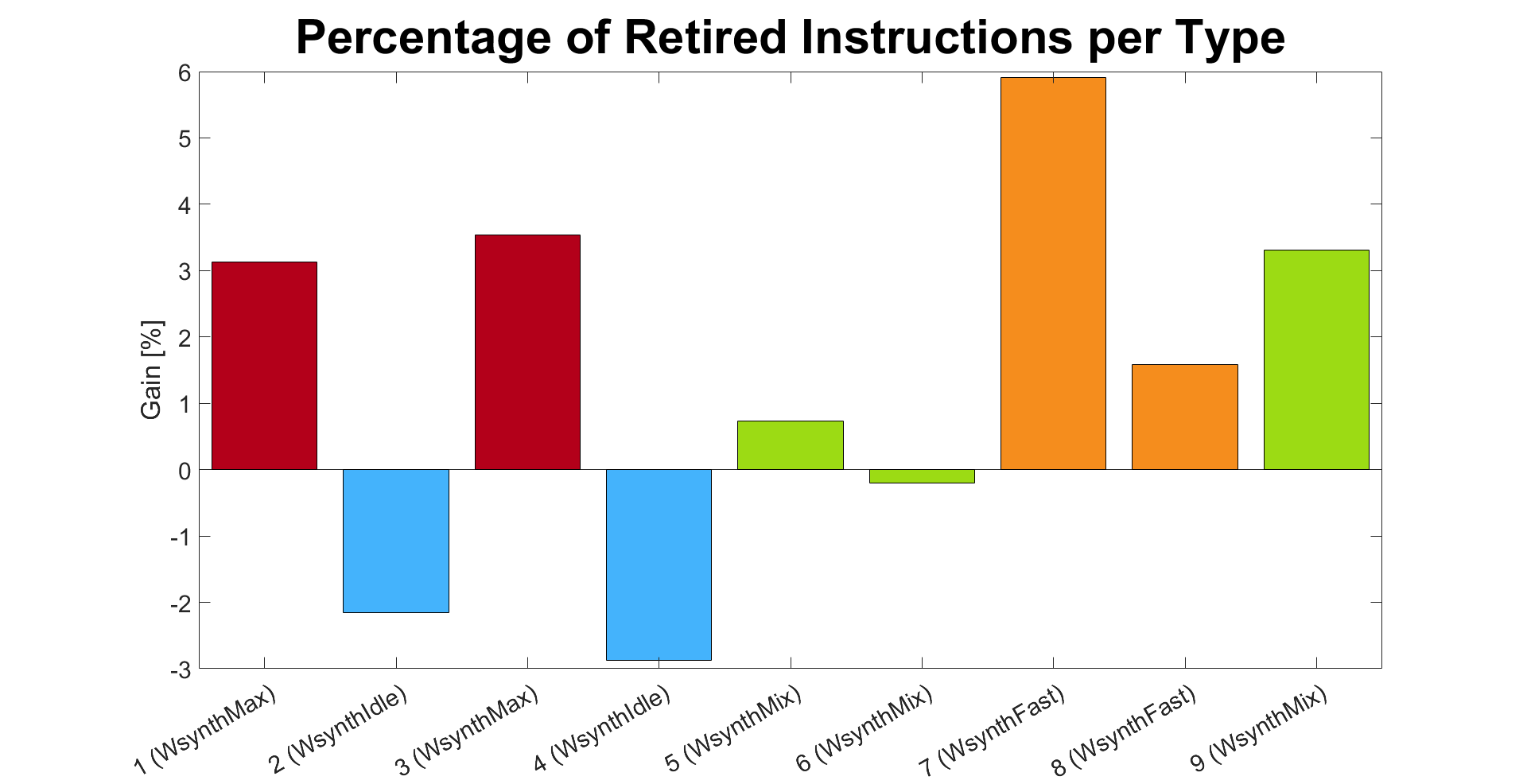}
	\caption{\label{fig:comp3_perf} Comparison between the \gls{pcf} control and the modified IBM OpenPOWER firmware with per-core temperature \gls{pid} for frequency reduction. The simulation time is $2s$}
\end{figure*}

Last, we compare the \gls{pcf} and the IBM control with per-core temperature \glspl{pid}. The \gls{pcf} control favors cores executing high-power instructions~\cite{gap8-hil} (\textit{core 1} and \textit{core 3} in this simulation with \texttt{Wsynth} benchmark), thus compensating the performance penalty showed in the previous test.
Results in Fig.~\ref{fig:comp3_perf} show a performance increase in executed instructions ranging from $+2.5\%$ to $+5\%$. This holds for cores with mixed instructions (up to $+3.5\%$) as well, while cores involved in less demanding workloads witness a decrease between $-2\%$ and $-3\%$. 
We conclude that the modified policy with per-core temperature \gls{pid} calculation can selectively boost the retired instructions, achieving a higher application performance on the \gls{hpc} chip while still meeting the thermal cap. 

\section{Conclusion}\label{sec:conclusions}

In this paper, we presented ControlPULP, the first \gls{hw}/\gls{sw} RISC-V control system for HPC processors that exploits multi-core capabilities to accelerate control algorithms, featuring a specialized \gls{dma} and fast interrupt handling and synchronization. 
We assess the \gls{hw}/\gls{sw} interface and take into account physical state variations surrounding the integrated controller by designing an agile, near-real-time closed-loop emulation framework that takes inspiration from the road paved by modern \gls{hesoc} platforms on \gls{fpga}. The framework relies on a power and thermal model of the controlled \gls{cpu} as plant, which is paired with workload instruction traces and the control algorithm to realize the closed-loop.
With the proposed multi-core architecture, a control policy step executes 4.9x faster than in a single-core configuration, increasing the hyper-period's \textit{slack} and thus enabling the implementation of more complex and computationally demanding control algorithms (for example, predictive policies) with fine-grained frequency dispatching.



\bmhead{Acknowledgments and funding}
This work was supported by the Spoke 1 on Future HPC of the Italian Research Center on High-Performance Computing, Big Data and Quantum Computing (ICSC) funded by MUR Mission 4 - Next Generation EU, by the European Project EuroHPC JU The European Pilot (g.a. 101034126), by EU H2020-JTI-EuroHPC-2019-1 project REGALE (g.n. 956560), by European Processor Initiative (EPI) SGA2 (g.a. 101036168), and by KDT TRISTAN project (g.a. 101095947). 

\bmhead{Author Contributions} Authors’ contributions are as follows: AO: system hardware design, investigation, experiments, writing—original draft. RB: system hardware design, text review. GB: system software design, text review. AV: system hardware design. MC: system hardware design. DR: text review. LB: text review and editing. AB: writing-review and editing.

\bmhead{Conflict of Interest} The authors have no competing interests as defined by Springer, or other interests that might be perceived to influence the results and/or discussion reported in this paper.







\bibliography{sn-bibliography}


\end{document}